\documentclass[conference,compsoc]{IEEEtran}

\usepackage{amsmath}
\usepackage{amssymb}
\usepackage{stmaryrd}
\usepackage{proof}
\usepackage{listings}
\usepackage{graphicx}
\usepackage{setspace}

\begin{document}

%%% our user-supplied commands

%%%
%%% spacing
%%%

\newlength{\indlen}
\setlength{\indlen}{8pt}
\newcommand{\ind}[1]{\hspace{#1\indlen}} % stands for indent
\newcommand{\figfill}[1][9pt]{\hspace{#1}\hfill}

\newcommand{\descitem}[1]{\item[]\hspace{-10pt}\textbf{#1}}

%%%
%%% theorem/proof constructs
%%%

\newtheorem{theorem}{Theorem}
\newtheorem{lemma}{Lemma}
\newtheorem{definition}{Definition}
\newtheorem{example}{Example}

% proof environment
% \iftechreport
%\newenvironment{proof}
%               {\hspace{-1\parindent}\textbf{Proof:}}
%               {\hfill $\square$}
% \else
% \newenvironment{myproof}
%                {\begin{IEEEproof}}
%                {\end{IEEEproof}}
% \fi

%%%
%%% for writing grammars and rules
%%%

\newcommand{\borspacing}{\hspace{.4em}}
\newcommand{\doublebar}{\protect\rule[-2pt]{1pt}{1em}\hspace{1pt}\protect\rule[-2pt]{1pt}{1em}}
\newcommand{\singlebar}{\protect\rule[-2pt]{1pt}{1em}}
\newcommand{\borstar}{\singlebar\borspacing}
\newcommand{\bor}{\borspacing\borstar} % stands for bar-or, for use in grammars

\newcommand{\rulename}[1]{\ensuremath{\text{\textsc{#1}}}}

%% commands related to sequences of syntactic objects

\newcommand{\seq}[1]{\vec{#1}}
\newcommand{\seqlen}[1]{\#(\seq{#1})}

%%%
%%% for defining commands that just set the font face
%%%

\newcommand{\definermcommand}[2]{\newcommand{#1}{\ensuremath{\mathrm{#2}}}}
\newcommand{\definesfcommand}[2]{\newcommand{#1}{\ensuremath{\mathsf{#2}}}}
\newcommand{\definebfcommand}[2]{\newcommand{#1}{\ensuremath{\mathbf{#2}}}}

%%%
%%% important names of things
%%%

\newcommand{\fadts}{\ensuremath{F^{\mathrm{ADT}}}}
\newcommand{\fadtsp}{\ensuremath{F^{\mathrm{ADT}+}}}

%%%
%%% set theory / meta-language notation and associated definitions
%%%

\newcommand{\powerset}{\ensuremath{\mathcal{P}}}
% \newcommand{\R}{\ensuremath{\mathbb{R}}}
% \newcommand{\Rp}{\ensuremath{\mathbb{R^{+}}}}
% \newcommand{\Rpinf}{\ensuremath{\mathbb{R^{+\infty}}}}
%\definebfcommand{\max}{max}
%\newcommand{\setapp}[2]{#1 \;@\; #2}

\definermcommand{\Dom}{Dom}
\definermcommand{\Ran}{Ran}

%% ordered tuples
\newcommand{\onetuple}[1]{\langle #1\rangle}
\newcommand{\pair}[2]{\langle #1 ,\; #2\rangle}
\newcommand{\pairsemi}[2]{\langle #1 ;\; #2\rangle}
\newcommand{\triple}[3]{\langle #1 ,\; #2, \; #3\rangle}
\newcommand{\quadtuple}[4]{\langle #1 ,\; #2, \; #3, \; #4\rangle}
\newcommand{\quintuple}[5]{\langle #1 ,\; #2, \; #3, \; #4, \; #5\rangle}
\newcommand{\sextuple}[6]{\langle #1 ,\; #2, \; #3, \; #4, \; #5, \; #6\rangle}
\newcommand{\tuple}[1]{\langle #1\rangle}

% array versions
\newcommand{\quadtuplearr}[4]{\langle\begin{array}[t]{@{}l@{}} #1 ,\; #2, \; #3, \; #4\rangle\end{array}}
\newcommand{\quintuplearr}[5]{\langle\begin{array}[t]{@{}l@{}} #1 ,\; #2, \; #3, \; #4, \; #5\rangle\end{array}}

% squashed versions
\newcommand{\triplesm}[3]{\langle #1 , #2, #3\rangle}
\newcommand{\quintuplesm}[5]{\langle #1 , #2,  #3,  #4,  #5\rangle}

\newcommand{\pairnl}[2]{\langle \begin{array}[t]{@{}l@{}}#1 ,\\ #2\rangle\end{array}}

%% set comprehension
\newcommand{\setcomprdivider}{\protect\rule[-2pt]{1pt}{1em}}
\newcommand{\setcompr}[2]{\{\, #1 \;\setcomprdivider\;\; #2 \,\}}
\newcommand{\setcomprarr}[2]{\{\, #1 \;\setcomprdivider\;\; \begin{array}[t]{@{}l@{}} #2 \,\}\end{array}}
\newcommand{\setcomprlong}[2]{\begin{array}[t]{@{}l@{}}\{\, #1 \\\;\;\;\;\setcomprdivider\;\; \begin{array}[t]{@{}l@{}}#2 \,\}\end{array}\end{array}}

%% relations
\newcommand{\field}[1]{\ensuremath{\mathrm{fld}(#1)}}

%% topology stuff
\newcommand{\ball}[2]{B_{#1}(#2)}
\newcommand{\cball}[2]{C_{#1}(#2)}

%% dependent function type
%\newcommand{\piabs}[3]{\Pi (#1\in #2).#3}

%%%
%%% PL stuff
%%%

\newcommand{\lamabs}[2]{\lambda #1.\,#2}
\newcommand{\lamabst}[3]{\lambda #1\!:\!#2.\,#3}
\newcommand{\lamabsnot}[2]{\lambda #1.\,#2}
\newcommand{\tplamabs}[2]{\Lambda #1.\,#2}
\newcommand{\lamrec}[3]{\lambda^{\mathrm{rec}}_{#1} #2.\,#3}
\newcommand{\tpapp}[2]{#1[#2]}
\newcommand{\foralltp}[2]{\forall #1.#2}
\newcommand{\kapp}[3]{#1(#2; #3)}
\newcommand{\kappnot}[2]{#1(#2)}
\definermcommand{\cl}{cl}

%% case expresions
\definebfcommand{\mycasename}{case}
\newcommand{\mycase}[1][\vec{c}]{\mycasename_{#1}}

% case as a functional constant
% \newcommand{\caseof}[3][\vec{c}]{\kapp{\case[#1]}{#2}{#3}}
% \newcommand{\caseofnot}[2][\vec{c}]{\kappnot{\case[#1]}{#2}}
% \newcommand{\caseofdef}[1]{\caseof{\vec{\tau}}{#1}}
%\newcommand{\caseoftypes}[4][\vec{c}]{\kapp{\case[#1]}{#2}{#3,#4}}
%\newcommand{\caseof}[3][\vec{c}]{\caseoftypes[#1]{\vec{\tau}}{#2}{#3}}

% case as it's own syntactic construct
\newcommand{\caseof}[3][\vec{c}]{\mycase[#1](#2;\;#3)}

% % a nicer way to write case when not in "PL mode"
\newcommand{\caseofpm}[2]{\mycase[](#1; #2)}
\newcommand{\pc}[2]{#1\mapsto #2}

% misc versions of case
%\newcommand{\caseof}[2]{\casename\;#1\;\mathbf{of}\;#2}
% \newcommand{\case}[3]{\ensuremath{\casename_{#1}^{#2\mapsto#3}}}
% \newcommand{\casedef}{\case{\vec{c}}{\vec{\tau}}{\tau}}
%\newcommand{\pc}[2]{#1\mapsto #2}

\definebfcommand{\ifname}{if}
\newcommand{\ifthenelse}[3]{\ifname\;#1\;\mathbf{then}\;#2\;\mathbf{else}\;#3}
\newcommand{\ifthenelsearr}[3]{\ifname\;#1\;\begin{array}[t]{@{}l@{}l@{}}\mathbf{then}\;#2\\ \mathbf{else}\;#3\end{array}}

\definebfcommand{\letname}{let}
\newcommand{\letin}[3]{\letname\;#1\;=\;#2\;\mathbf{in}\;#3}

\newcommand{\ctxelem}[2]{#1:#2}

%% typing
\newcommand{\wfj}[2][\Gamma]{#1\vdash #2}
\newcommand{\typejprot}[3]{#1 \vdash #2 : #3}
\newcommand{\typej}[3][\Gamma]{\typejprot{#1}{#2}{#3}}
\newcommand{\kindj}[2][\Gamma]{#1 \vdash #2 : *}
\newcommand{\ctxtp}[2]{(#1\vdash #2)}
\definermcommand{\gr}{gr}
\definermcommand{\ctors}{ctors}
\definermcommand{\ctp}{ctp}
\definermcommand{\ar}{ar}

%% term contexts
\newcommand{\tctx}[2]{\ensuremath{#1\{#2\}}}
\newcommand{\tctxemp}{\{\}}
\newcommand{\Cany}{C^*}

%% evaluation
\newcommand{\rrto}{\longrightarrow}
\newcommand{\rrtostar}{\longrightarrow^{*}}
\definermcommand{\NF}{NF}
\newcommand{\term}[1]{#1\!\downarrow}
\newcommand{\termat}[2]{#1\!\downarrow\! #2}
\newcommand{\nonterm}[1]{#1\!\uparrow}

%% contextual & substitution equivalence
\newcommand{\ctxeq}{\cong}

\newcommand{\expset}[2][\Gamma]{\llbracket #1\vdash #2\rrbracket}
\newcommand{\expsete}[1]{\expset[\Gammae]{#1}}
\newcommand{\expseta}[1]{\expset[\Gammaa]{#1}}
\newcommand{\expsetq}[1]{\expset[\Gammaq]{#1}}
\newcommand{\expsetqe}[1]{\expset[\Gammaq,\Gammae]{#1}}

%% examples

% fix
\definebfcommand{\fixname}{fix}
\newcommand{\fix}{\fixname}
\newcommand{\fixapp}[2][\tau]{\tpapp{\fix}{#1}(#2)}
\newcommand{\fixappnot}[1]{\fix(#1)}
\newcommand{\fixappnop}[2][\tau]{\tpapp{\fix}{#1}\;#2}
\newcommand{\fixappnotp}[1]{\fix\;#1}
% \newcommand{\fixapp}[2][\tau]{\kapp{\fix}{#1}{#2}}
% \newcommand{\fixappnot}[1]{\kappnot{\fix}{#1}}
%\newcommand{\fixapp}[1]{\kappnot{\fix}{#1}}
%\newcommand{\fixname}[1][x]{\ensuremath{\mathbf{fix}_{#1}}}
%\newcommand{\fix}[2][x]{\fixname[#1]\;#2}

% Rec
\definesfcommand{\Rec}{Rec}
\definesfcommand{\roll}{roll}

% polymorphic identity
\definesfcommand{\id}{id}

% unit (final) and empty (initial) types
\newcommand{\Empty}{\ensuremath{\mathsf{Emp}}}
\newcommand{\Unit}{\ensuremath{\mathsf{Unit}}}
\newcommand{\unit}{\ensuremath{\mathsf{unit}}}
\newcommand{\QUnit}{\Q_{\Unit}}

% booleans
\definesfcommand{\Bool}{B}
\definesfcommand{\true}{true}
\definesfcommand{\false}{false}

% naturals
%\newcommand{\N}{\ensuremath{\mathsf{N}}}
\definesfcommand{\N}{N}
\newcommand{\zeroN}{0_\N}
\newcommand{\oneN}{1_{\N}}
\newcommand{\plusN}{+_\N}
\newcommand{\timesN}{*_{\N}}
\newcommand{\divN}{/_{\N}}
\newcommand{\Z}{\ensuremath{\mathsf{Z}}}
\newcommand{\Ninf}{\ensuremath{\mathsf{N}^{\infty}}}

% reals
\newcommand{\R}{\ensuremath{\mathsf{R}}}
\newcommand{\Rp}{\ensuremath{\mathsf{R}^{+}}}
\newcommand{\Rpinf}{\ensuremath{\mathsf{R^{+\infty}}}}
\newcommand{\plusR}{+_\R}
\newcommand{\zeroR}{0_\R}
\newcommand{\oneR}{1_{\R}}
\newcommand{\timesR}{*_{\R}}
\newcommand{\leqR}{\leq_\R}
\newcommand{\geqR}{\geq_\R}
\newcommand{\negR}{-_\R}
\newcommand{\dR}{d_\R}
%\newcommand{\Qrat}{\ensuremath{\mathsf{Q}}}
%\definesfcommand{\PRat}{PRat}
%\definesfcommand{\Rat}{Rat}

% floating-point
\definesfcommand{\Float}{Float}
%\definesfcommand{\Float}{Fl}
\newcommand{\plusFl}{+_{\Float}}
\newcommand{\zeroFl}{0_{\Float}}

% fixed-point
%\definesfcommand{\Fixed}{Fixed}
\definesfcommand{\Fixed}{Fx}
\newcommand{\plusFx}{+_{\Fixed}}
\newcommand{\zeroFx}{0_{\Fixed}}
\newcommand{\plusFxq}{+_{\Fixed}^{\mathrm{q}}}
\newcommand{\timesFx}{*_{\Fixed}}
\newcommand{\timesFxq}{*_{\Fixed}^{\mathrm{q}}}
\newcommand{\negFx}{-_\Fixed}

% Int type
\definesfcommand{\Int}{int}

% option type
\definesfcommand{\Opt}{Opt}
\definesfcommand{\None}{None}
\definesfcommand{\Some}{Some}

% list type
\definesfcommand{\List}{List}
\definesfcommand{\nil}{nil}
% \definesfcommand{\cons}{cons}
%\newcommand{\nil}{[]}
%\newcommand{\consname}{::}
%\newcommand{\cons}[2]{#1 \consname #2}
\definesfcommand{\consname}{cons}
\newcommand{\cons}[2]{\consname\;#1\;#2}

\definesfcommand{\head}{head}
\definesfcommand{\tail}{tail}
\newcommand{\mklist}[1]{[#1]}
\definesfcommand{\len}{len}
\newcommand{\nth}[2]{#1.[#2]}
\newcommand{\listinj}[2]{#1\leftrightarrow#2}
\definesfcommand{\foldr}{foldr}

% partiality monad
\definesfcommand{\Par}{Par}
\definesfcommand{\ret}{ret}
\definesfcommand{\bind}{bind}
%\definesfcommand{\fix}{fix}

%%%
%%% for writing approximations
%%%

%% higher-order logic
\definesfcommand{\Prop}{Prop}
\definesfcommand{\True}{True}
\definesfcommand{\False}{False}
\newcommand{\truej}[1]{\vDash #1 \text{ \textbf{true}}}
\newcommand{\entjprot}[2]{#1 \vdash #2 \text{ \textbf{true}}}
\newcommand{\entj}[2][\cctxdef]{\entjprot{#1}{#2}}
\newcommand{\cctx}[2]{\pairsemi{#1}{#2}}
\newcommand{\cctxdef}[1][]{\cctx{\Gamma_{#1}}{\phi_{#1}}}
\newcommand{\ssatj}[2]{#1\vDash #2}
\newcommand{\typejcctx}[3]{\typejprot{#1}{#2}{#3}}
\newcommand{\typejcctxdef}[3]{\typejprot{\cctxdef[#1]}{#2}{\cctxdef[#3]}}
\newcommand{\repr}[1]{\ulcorner #1 \urcorner}

\definebfcommand{\isapprox}{isapprox}
\definebfcommand{\approxtp}{approx}

\newcommand{\cctxapprprot}[1]{\cctx{\eaqActx[#1]}{\phictx[#1]}}
\newcommand{\cctxappr}[1][\apprctx]{\cctxapprprot{#1}}
\newcommand{\ssatjapprctx}[2][\apprctx]{\ssatj{#2}{\cctxappr[#1]}}

\newcommand{\lammeta}[2]{\bar{\lambda} #1.\,#2}
\newcommand{\metactx}[2]{#1;#2}
\newcommand{\wfpred}[2][\metactx{\vec{\xi}}{\Gamma}]{#1\vdash #2}
\newcommand{\typejpred}[2][\apprctx]{#1\vdash #2}
\newcommand{\oldtruejprot}[2]{\vdash #2(#1) \text{ \textbf{true}}}
\newcommand{\oldtruej}[2][\Theta]{\oldtruejprot{#1}{#2}}
\newcommand{\truepred}{\top}
\newcommand{\sigmaAs}{\pairsemi{\sigma}{\vec{\A}}}
\newcommand{\truejsigmaAs}[1]{\oldtruej[\sigmaAs]{#1}}

%% quantification types
\newcommand{\quantt}[4]{\quadtuple{#1}{#2}{#3}{#4}}
\newcommand{\quanttctx}[5]{\quintuple{#1}{#2}{#3}{#4}{#5}}
\newcommand{\Q}{\ensuremath{\mathcal{Q}}}
\newcommand{\QQ}[1][\Q]{Q_{#1}}
\newcommand{\leqQ}[1][\Q]{\leq_{#1}}
\newcommand{\geqQ}[1][\Q]{\geq_{#1}}
\newcommand{\gtQ}[1][\Q]{>_{#1}}
\newcommand{\plusQ}[1][\Q]{+_{#1}}
\newcommand{\zeroQ}[1][\Q]{0_{#1}}
\newcommand{\inftyQ}[1][\Q]{\infty_{#1}}
\newcommand{\ltQ}[1][\Q]{<_{#1}}
%\definesfcommand{\downname}{down}
\newcommand{\downname}{\downarrow}
\newcommand{\downnameQ}[1][\Q]{\downname_{#1}}
\newcommand{\downQ}[3][\Q]{#2\!\downnameQ[#1]\!#3}
\newcommand{\IQ}[1][\Q]{I_{#1}}

%% useful contexts
\newcommand{\Gammae}{\Gamma^{\mathrm{e}}}
\newcommand{\Gammaa}{\Gamma^{\mathrm{a}}}
\newcommand{\Gammaq}{\Gamma^{\mathrm{q}}}
\newcommand{\GammaA}{\Gamma^{\mathrm{A}}}

\newcommand{\QR}{\Q_{\R}}
\newcommand{\qto}{\Rightarrow}

%% approximate equality
\newcommand{\aeqrel}[1][]{\approx_{#1}}
\newcommand{\aeq}[4][]{#3\aeqrel[#1]^{#2}\!#4}

\newcommand{\homeq}[1]{\mathsf{All(#1)}}
\newcommand{\funeq}[1]{\mathsf{DistFun}(#1)}

%% approximations themselves
\newcommand{\apprtuple}[5]{\quintuple{#1}{#2}{#3}{#4}{#5}}
\newcommand{\apprtuplesm}[5]{\quintuplesm{#1}{#2}{#3}{#4}{#5}}
\newcommand{\llbrace}{\{\hspace{-2.1pt}|}
\newcommand{\rrbrace}{|\hspace{-2.1pt}\}}
\newcommand{\apprprot}[3]{\llbrace#3\rrbrace_{#1}^{#2}}
\newcommand{\appr}[3][]{\apprprot{#1}{#2}{#3}}

\newcommand{\A}{\ensuremath{\mathcal{A}}}
\newcommand{\norm}[1]{\|#1\|}

\newcommand{\EA}[1][]{E_{\A_{#1}}}
\renewcommand{\AA}[1][]{A_{\A_{#1}}}

\newcommand{\QA}[1][]{\Q_{\A_{#1}}}
\newcommand{\QQA}[1][]{Q_{\A_{#1}}}
\newcommand{\leqA}[1][\A]{\leq_{#1}}
\newcommand{\ltA}[1][]{<_{\A_{#1}}}
\newcommand{\geqA}[1][]{\geq_{\A_{#1}}}
\newcommand{\plusA}[1][\A]{+_{#1}}
\newcommand{\zeroA}[1][\A]{0_{#1}}
\newcommand{\downnameA}[1][]{\downnameQ[#1]}
\newcommand{\downA}[3][]{\downQ[#1]{#2}{#3}}
\newcommand{\IA}[1][]{I_{\A_{#1}}}

%% approximation functions
\newcommand{\Aapp}[2]{#1(#2)}

%% polymorphic approximations
%\newcommand{\F}{\ensuremath{\mathbb{F}}}
\newcommand{\F}{\ensuremath{\mathcal{F}}}
\newcommand{\QF}[1][]{\Q_{\F_{#1}}}
\newcommand{\QQF}[1][]{Q_{\F_{#1}}}
\newcommand{\leqF}[1][]{\leq_{\F_{#1}}}
\newcommand{\ltF}[1][]{<_{\F_{#1}}}
\newcommand{\geqF}[1][]{\geq_{\F_{#1}}}
\newcommand{\plusF}[1][]{+_{\F_{#1}}}
\newcommand{\zeroF}[1][]{0_{\F_{#1}}}
\newcommand{\EF}[1][]{E_{\F_{#1}}}
\newcommand{\AF}[1][]{A_{\F_{#1}}}

% primed versions of some of the above
\newcommand{\EFp}[1][]{E_{\F'_{#1}}}
\newcommand{\AFp}[1][]{A_{\F'_{#1}}}
\newcommand{\QQFp}[1][]{Q_{\F'_{#1}}}
\newcommand{\plusFp}[1][]{+_{\F'_{#1}}}
\newcommand{\zeroFp}[1][]{0_{\F'_{#1}}}

%% approximation contexts
\newcommand{\apprctx}{\Xi}

\newcommand{\aelem}[4]{(#1,#2,#3)\!:\!#4}
\newcommand{\aelemvar}[2]{\aelem{#1^{\mathrm{e}}}{#1^{\mathrm{a}}}{#1^{\mathrm{q}}}{#2}}
\newcommand{\xq}[1][]{x^{\mathrm{q}}_{#1}}
\newcommand{\xa}[1][]{x^{\mathrm{a}}_{#1}}
\newcommand{\xe}[1][]{x^{\mathrm{e}}_{#1}}
\newcommand{\aelemdef}[1][]{\aelem{\xe[#1]}{\xa[#1]}{\xq[#1]}{\F_{#1}}}
\newcommand{\aelemA}[1]{\aelem{\xe}{\xa}{\xq}{\A_{#1}}}
\newcommand{\aelemF}[1]{\aelem{\xe}{\xa}{\xq}{\F_{#1}}}
\newcommand{\aelemabbr}[2]{#1:#2}
\newcommand{\aelemmulti}[2]{\aelemabbr{\vec{#1}}{\vec{#2}}}

\newcommand{\yq}[1][]{y^{\mathrm{q}}_{#1}}
\newcommand{\ya}[1][]{y^{\mathrm{a}}_{#1}}
\newcommand{\ye}[1][]{y^{\mathrm{e}}_{#1}}

\newcommand{\zq}[1][]{z^{\mathrm{q}}_{#1}}
\newcommand{\za}[1][]{z^{\mathrm{a}}_{#1}}
\newcommand{\ze}[1][]{z^{\mathrm{e}}_{#1}}

\newcommand{\bq}[1][]{b^{\mathrm{q}}_{#1}}
\newcommand{\ba}[1][]{b^{\mathrm{a}}_{#1}}
\newcommand{\be}[1][]{b^{\mathrm{e}}_{#1}}

% README: \le, \la, and \lq are existing (and useful!) commands...
\newcommand{\llq}[1][]{l^{\mathrm{q}}_{#1}}
\newcommand{\lla}[1][]{l^{\mathrm{a}}_{#1}}
\newcommand{\lle}[1][]{l^{\mathrm{e}}_{#1}}

\newcommand{\atelem}[4]{(#1,#2,#3):#4}
\newcommand{\Xq}[1][]{X^{\mathrm{q}}_{#1}}
\newcommand{\Xa}[1][]{X^{\mathrm{a}}_{#1}}
\newcommand{\Xe}[1][]{X^{\mathrm{e}}_{#1}}
\newcommand{\zp}[1][]{z^{+}_{#1}}
\newcommand{\zz}[1][]{z^{0}_{#1}}
\newcommand{\atelemdef}[1][]{\atelem{\Xe[#1]}{\Xa[#1]}{\Xq[#1]}{\xi_{#1}}}
\newcommand{\atelemabbr}[2]{#1:#2}
\newcommand{\atzelemabbr}[3]{#1,#2:#3}

\newcommand{\phielem}[2]{#1.(#2)}
\newcommand{\phielemdef}{\phielem{\Gamma}{\phi}}

\newcommand{\ectx}[1][\apprctx]{|#1|^{\mathrm{e}}}
\newcommand{\actx}[1][\apprctx]{|#1|^{\mathrm{a}}}
\newcommand{\qctx}[1][\apprctx]{|#1|^{\mathrm{q}}}
\newcommand{\eqctx}[1][\apprctx]{|#1|^{\mathrm{eq}}}
\newcommand{\qActx}[1][\apprctx]{|#1|^{\mathrm{qA}}}
\newcommand{\eaqActx}[1][\apprctx]{|#1|^{\mathrm{eaqA}}}
\newcommand{\Actx}[1][\apprctx]{|#1|^{\mathrm{A}}}
\newcommand{\phictx}[1][\apprctx]{|#1|^{\mathrm{p}}}

% typing judgments related to approximation contexts
\newcommand{\wfappr}[1]{\vdash #1}
\newcommand{\typejF}[2][\apprctx]{#1\vdash #2}
\newcommand{\typejT}[3][\apprctx]{#1\vdash #2:#3}

%% extended expressions and approximation valuations
\newcommand{\Exp}{\Upsilon}
\newcommand{\avalsat}[2][\apprctx]{#2\models #1}

\newcommand{\apprval}[2]{\pairsemi{#1}{#2}}
\newcommand{\apprvaldef}[1][]{\apprval{\Theta_{#1}}{\sigma_{#1}}}
\newcommand{\avalj}[2][\apprctx]{#1\vdash#2}
\newcommand{\avaljdef}{\avalj{\apprvaldef}}

%% building approximations
\newcommand{\approxname}[1]{\ensuremath{\mathbf{#1}}}

% fixed-point and floating-point
%\definesfcommand{\FloatA}{Fl}
%\definesfcommand{\FixedA}{Fx}
%\newcommand{\FixedA}{\A_{\mathsf{Fx}}}
%\newcommand{\FixedA}{\mathcal{Fx}}
%\newcommand{\FixedA}{\mathscr{F}}
\newcommand{\FixedA}{\approxname{Fx}}
\newcommand{\FloatA}{\approxname{Fl}}
\newcommand{\NatA}{\approxname{Nat}}
\definesfcommand{\ckover}{ck\text{-}over}

% rounding approximation
\newcommand{\TruncA}[1][f]{\approxname{Trunc}(#1)}

% sampling approximation
% \newcommand{\SampleAN ame}{\approxname{RSmpl}}
% \newcommand{\SampleA}[1]{\SampleAName(#1)}
\newcommand{\SampleFxA}[1][f]{\approxname{SmplFx}(#1)}
\definesfcommand{\inter}{inter}

% context-related approximations
\newcommand{\CtxAprot}[2]{#1\Vdash #2}
\newcommand{\CtxA}[2][\apprctx]{\CtxAprot{#1}{#2}}

% exact and hom approximations
\newcommand{\ExactA}[1]{\approxname{Exact}(#1)}
\newcommand{\ExactStrA}[1]{\approxname{ExS}(#1)}
\newcommand{\HomA}[1]{\approxname{Hom}(#1)}
%\newcommand{\EmptyA}[1]{\mathsf{Emp}(#1)}
%\newcommand{\SetA}[2]{\mathsf{Set}(#1,#2)}

% list approximations
\newcommand{\Lone}{L_1}
\newcommand{\Linf}{L_\infty}
\newcommand{\LoneA}[1]{\approxname{L}_1(#1)}
\newcommand{\LinfA}[1]{\approxname{L}_\infty(#1)}

% functional-ish approximations
%\newcommand{\ato}{\Rightarrow}
\newcommand{\apprto}{\Rightarrow}
\newcommand{\forallA}[2]{\forall #1.#2}

\newcommand{\apprimp}{\Rightarrow}

\newcommand{\piappr}[2]{\Pi(#1).#2}
\newcommand{\piapprdef}[2][]{\piappr{\aelemvar{x_{#1}}{\F_{#1}}}{#2}}
\newcommand{\piapprtdef}[2][]{\piappr{\atzelemabbr{X_{#1}}{z_{#1}}{\xi_{#1}}}{#2}}

\newcommand{\piapprmulti}[2][]{\piappr{\aelemmulti{x}{\F_{#1}}}{#2}}
\newcommand{\piapprmultiy}[2][]{\piappr{\aelemmulti{y}{\F_{#1}}}{#2}}
\newcommand{\piapprmultiA}[1]{\piappr{\aelemabbr{\vec{x}}{\vec{\A}}}{#1}}
\newcommand{\piapprtmulti}[1]{\piappr{\atzelemabbr{\vec{X}}{\vec{z}}{\vec{\xi}}}{#1}}

\newcommand{\taue}{\tau^{\mathrm{e}}}
\newcommand{\taua}{\tau^{\mathrm{a}}}

%% Dedekind cuts
\newcommand{\Ded}[1]{\mathrm{Ded}(#1)}
\newcommand{\close}[1]{\mathrm{Cl}(#1)}

\newcommand{\DeA}[1][]{D^{\mathrm{e}}_{\A_{#1}}}
\newcommand{\DaA}[1][]{D^{\mathrm{a}}_{\A_{#1}}}

%% floating-point numbers
%\newcommand{\Float}{\ensuremath{\mathbb{F}_{64}}}
%\newcommand{\roundF}{\round_{\Float}}
%\newcommand{\NaN}{\ensuremath{\mathrm{NaN}}}
%\newcommand{\dF}{\ensuremath{d_{64}}}

%% Taylor expansion
\definesfcommand{\Taylor}{Taylor}

%%%
%%% compilation rules
%%%

\newcommand{\compilestonoerr}[3][\apprctx]{#1\vdash #2 \leadsto #3}
\newcommand{\compilesto}[5][\apprctx]{#1\vdash #2 \leadsto #3 \leq #4 : #5}
\newcommand{\compilestonl}[5][\apprctx]{#1\vdash \begin{array}[t]{@{}l@{}}#2 \leadsto #3\\ \leq \begin{array}[t]{@{}l@{}}#4 : #5\end{array}\end{array}}
\newcommand{\compilestonla}[5][\apprctx]{\begin{array}[t]{@{}l@{}}#1\vdash #2 \leadsto #3 \leq #4 \\\hspace{20pt} : #5\end{array}}
\newcommand{\compilestonle}[5][\apprctx]{\begin{array}[b]{@{}l@{}}#1\vdash #2 \\\hfill\leadsto #3 \leq #4 : #5\end{array}}
\newcommand{\K}{\ensuremath{\mathcal{K}}}
\newcommand{\compilestonlnl}[5][\apprctx]{#1\;\begin{array}[t]{@{}l@{}}\vdash #2 \\\leadsto #3 \\\leq #4 : #5\end{array}}
%\newcommand{\compilestonlarr}[5][\apprctx]{\begin{array}[b]{@{}l@{}}#1\vdash #2 \leadsto #3\\\hspace{15pt} \leq #4 : #5\end{array}}

% for the case rule
\newcommand{\apprctxexp}{\apprctx^{\mathrm{x}}}
\newcommand{\matches}[5]{\mathbf{matches}(#1,#2\mapsto #3 \wedge #4\leq #5)}
\definebfcommand{\matchboundname}{match\text{-}dist\text{-}bound}
\newcommand{\matchbound}[8]{\matchboundname(#1\mapsto #2\leftrightarrow #3\mapsto#4 \leq #5 \Rightarrow #6\leftrightarrow #7\leq #8)}
\newcommand{\matchboundnl}[8]{\matchboundname(\begin{array}[t]{@{}l@{}}#1\mapsto #2\leftrightarrow #3\mapsto#4\\ \leq #5 \Rightarrow #6\leftrightarrow #7\leq #8)\end{array}}

% for writing examples
\newcommand{\compilestoeg}[4]{#1 \leadsto #2 \leq #3 : #4}

%%%
%%% approximation examples
%%%

% PID example
\definesfcommand{\PID}{PID}
\definesfcommand{\PIDStep}{PIDStep}

\definesfcommand{\PlusErr}{PlusErr}
\definesfcommand{\TimesErr}{TimesErr}
\definesfcommand{\IntErr}{IntErr}
%\definesfcommand{\DerivErr}{DerivErr}
\definesfcommand{\DerivErr}{DErr}

% sum
%\definesfcommand{\sumr}{sum^{\R}}
\definesfcommand{\sumr}{sum}
\definesfcommand{\sumfl}{sum^{\Float}}
\definesfcommand{\sumfx}{sum^{\Fixed}}
\definesfcommand{\sumerr}{sum^{err}}

\newcommand{\ckoversub}[1]{\mathsf{ck\text{-}over}_{#1}}
\newcommand{\ckoversum}{\ckoversub{}}
%\newcommand{\ckoversum}{\ckover{\sumr}}
%\definesfcommand{\ckoversum}{ck\text{-}over\text{-}sum}
\newcommand{\econs}{e_{\consname}}
\newcommand{\acons}{a_{\consname}}
\newcommand{\qcons}{q_{\consname}}
\newcommand{\phinil}{\phi_{\nil}}
\newcommand{\phicons}{\phi_{\consname}}
\newcommand{\phiside}{\phi_{\mathrm{side}}}
\newcommand{\phifix}{\phi_{\fixname}}

% product
\definesfcommand{\prodr}{prod^{\R}}
\definesfcommand{\prodfx}{prod^{\Fixed}}
\newcommand{\ckoverprod}{\ckoversub{\prodr}}
%\definesfcommand{\ckoverprod}{ck\text{-}over\text{-}sum}
\definesfcommand{\maxprod}{max\text{-}prod}

% abs
%\definesfcommand{\absr}{abs^\R}
\definesfcommand{\absr}{abs}
\definesfcommand{\absfx}{abs^\Fixed}

% squaring
%\definesfcommand{\sqr}{sq^\R}
\definesfcommand{\sqr}{sq}
\definesfcommand{\sqfx}{sq^\Fixed}

% power
\definesfcommand{\powr}{pow^\R}
\definesfcommand{\powfx}{pow^\Fixed}

% maximums and upper bounds for approx's
\newcommand{\pimaxfun}[3]{\piappr{#1\!:\!\mathbf{max}(#2)}{#3}}
\newcommand{\ub}[1][\F]{\mathsf{ub}_{#1}}
\newcommand{\piub}[3]{\piappr{#1\!:\!\mathbf{ub}(#2)}{#3}}

% error functions that distribute over list
\newcommand{\pimaperr}[3]{\piappr{#1\!:\!\mathbf{maperr}(#2)}{#3}}
\definebfcommand{\folderrname}{folderr}
\newcommand{\pifolderr}[3]{\piappr{#1\!:\!\folderrname(#2)}{#3}}

\newcommand{\nne}{n^{\mathrm{e}}}
\newcommand{\nna}{n^{\mathrm{a}}}
\newcommand{\nnq}{n^{\mathrm{q}}}

\newcommand{\limA}[1][\A]{\mathsf{lim}^{#1}}

\newcommand{\Lperf}{L_{\mathrm{perf}}}
\newcommand{\LperfA}[1]{L_{\mathrm{perf}}(#1)}
\definesfcommand{\perf}{perf}
\newcommand{\perferr}[1][\F]{\mathsf{perf\text{-}err}^{#1}}
\definesfcommand{\map}{map}
\definesfcommand{\reduce}{reduce}
\definesfcommand{\redseq}{red\text{-}seq}

\definesfcommand{\folderrinf}{fold\text{-}err_{\infty}}

%%% Local Variables: 
%%% mode: latex
%%% TeX-master: "approx-compiler"
%%% End: 

%%% options for listings
\lstset{ %
  language=Haskell,
  basicstyle=\ttfamily\footnotesize,
  keywordstyle=\bf,
  showstringspaces=false,
  captionpos=b,
  mathescape,
  escapeinside={(*@}{@*)}
}

%%%
%%% Header information
%%%

%% IEEE version for LICS

\title{A Semantics for Approximate Program Transformations}

\author{\IEEEauthorblockN{Edwin Westbrook and Swarat Chaudhuri}
\IEEEauthorblockA{Department of Computer Science, Rice University\\
Houston, TX 77005\\
Email: \{emw4,swarat\}@rice.edu}}

\maketitle

%% LNCS version

% \title{A Semantics for Approximate Program Transformations}
% \subtitle{Subtitle Text, if any}

% \author{Edwin Westbrook \and Swarat Chaudhuri}
% \institute{Rice University, Houston, TX 77005, USA\\
% \email{\{emw4, swarat\}@rice.edu}
% }

% \maketitle

%% Old, ACM version

% \conferenceinfo{POPL '13}{January 23--25, Rome, Italy.} 
% \copyrightyear{2013}
% \copyrightdata{[to be supplied]} 

% %\titlebanner{banner above paper title}        % These are ignored unless
% %\preprintfooter{short description of paper}   % 'preprint' option specified.

% \title{A Semantics for Approximate Program Transformations}
% %\subtitle{Subtitle Text, if any}

% \authorinfo{Edwin Westbrook \and Swarat Chaudhuri}
%            {Rice University}
%            {\{emw4,swarat\}@rice.edu}
% % \authorinfo{Name2\and Name3}
% %            {Affiliation2/3}
% %            {Email2/3}

% \maketitle

%%%
%%% Abstract
%%%

\begin{abstract}
  An \emph{approximate program transformation} is a transformation
  that can change the semantics of a program within a specified
  empirical error bound. Such transformations have wide applications:
  they can decrease computation time, power consumption, and memory
  usage, and can, in some cases, allow implementations of incomputable
  operations. Correctness proofs of approximate program
  transformations are by definition quantitative. Unfortunately,
  unlike with standard program transformations, there is as of yet no
  modular way to prove correctness of an approximate transformation
  itself. Error bounds must be proved for each transformed program
  individually, and must be re-proved each time a program is modified
  or a different set of approximations are applied. 

  In this paper, we give a semantics that enables quantitative
  reasoning about a large class of approximate program transformations
  in a local, composable way. Our semantics is based on a notion of
  distance between programs that defines what it means for an
  approximate transformation to be correct up to an error bound.  The
  key insight is that distances between programs cannot in general be
  formulated in terms of metric spaces and real numbers. Instead, our
  semantics admits natural notions of distance for each type
  construct; for example, numbers are used as distances for numerical
  data, functions are used as distances for functional data, and
  polymorphic lambda-terms are used as distances for polymorphic
  data. We then show how our semantics applies to two example
  approximations: replacing reals with floating-point numbers, and
  loop perforation.
\end{abstract}

% \category{CR-number}{subcategory}{third-level}

% \terms
% term1, term2

% \keywords
% keyword1, keyword2

%%%%%%%%%%%%%%%%%%%%%%%%%%%%%%%
%%%%%       Section       %%%%%
%%%%%%%%%%%%%%%%%%%%%%%%%%%%%%%
\section{Introduction}

Approximation is a fundamental concept in engineering and computer
science, including notions such as floating-point
numbers, lossy compression, and approximation algorithms for NP-hard
problems. Such techniques are often used to trade off accuracy of the
result for \textbf{reduced resource usage}, for resources such as
computation time, power, and memory. In addition, some approximation
techniques are also used to ensure \textbf{computability}. For
example, true representations of real numbers (e.g.,
\cite{edalat05,boehm86}), require some operations, such as comparison,
to be incomputable; floating-point comparison, in contrast,
is efficiently decidable on modern computers.

Recently, there has been a growing interest in \emph{language-based
  approximations}, where \emph{approximate program transformations}
are performed by the programming language environment
\cite{zhu12,misailovic11,sidiroglou11,sampson11,chaudhuri11,chaudhuri10,reed10}.
Such approaches allow the user to give an \emph{exact program} as a
specification, and then apply some set of transformations to this
specification, yielding an \emph{approximate program}. The goal is for
approximations to be performed on behalf of the programmer, either
fully automatically or with only high-level input from the user, while
still maintaining a given error bound; i.e., the goals are
\textbf{automation} and \textbf{correctness}.  These goals can in turn
increase programmer productivity while helping to remove programmer
errors, where the latter can be especially important, for example, in
safety-critical systems.

This leads us to a fundamental question: what does it mean for an
approximate transformation to be correct? A good answer to this
question must surely be quantitative, since approximate
transformations should not change the output by too much; i.e., they
must respect user-specified error bounds. Correctness should also be
modular, meaning that,
%: (1) When an approximate transformation $T$ is
%applied to programs $P_1,\dots, P_k$, we should be able to reuse as
 in settings where approximate transformations
$T_1, \dots, T_k$ are applied together to a program $P$, it should be possible to
reduce the proof of correctness of $\{T_1, \dots, T_k\}$ to individual proof
obligations for the $T_i$s.
%, along with the obligation to prove
%some reasonable side conditions. 
Current formal approaches to approximate program transformations,
however, do not permit such modular reasoning about approximations.
Typically, they are tailored to specific forms of approximation --- for
example, the use of floating-point numbers or loop perforation~\cite{sidiroglou11} (skipping certain
iterations in long-running loops). Even when multiple approximations
can be combined, reasoning about them is monolithic; in the above
example, if $T_1$ is changed slightly, we would need to re-prove the
correctness of $P$ with respect to not just $T_1$ but $\{T_1, \dots,
T_k\}$.  In addition, current approaches to approximate
computation~\cite{carbin12,zhu12,sidiroglou11,sampson11,chaudhuri11} are almost universally restricted to
first-order programs over restricted data types.
% // Swarat: I do not think even Reed and Pierce would claim that
% // their paper is about approximate computation.
% // Also, the continuity analysis paper does not talk about
% approximate computation
%, with the
%notable exception of Reed and Pierce \cite{reed10}.

% This leads us to a fundamental question: what does it mean for an
% approximate transformation to be correct? A good answer to this
% question must surely be quantitative, since approximate
% transformations should not change the output by too much; i.e., they
% must respect user-specified error bounds. Correctness should also be
% modular, meaning that it should be possible to prove a quantitative
% error bound for each approximate transformation individually,
% independent of the program to which they are applied or of each other.
% Current approaches to correctness, however, require the whole program
% being transformed must be analyzed or tested, to compare the results
% after approximate transformations to those before. Thus, not only is
% there no general way to prove an approximate transformation correct,
% but there is no way to talk about how much error each individual
% transformation contributes to the overall error of an approximate
% program.

In this work, we improve on this state of the art by giving a general,
composable semantics for program approximation, for higher-order
programs with polymorphic types. In our semantics, individual 
approximate transformations are proved to be quantifiably
correct, i.e., to induce a given local error expression. Error
expressions are then combined compositionally, yielding a top-level
error expression for the whole program that is built up from the
errors of the individual approximate transforms being used.  This
approach has a number of benefits. First, it allows for more {\em
  portable} proofs:
% approximate
%transformations can be analyzed and proved correct on their own,
%without reference to the programs in which they are being used.  Thus
an 
approximate transformation $T$ can now be proved correct once, and the
resulting error expression can be used many times in many
different contexts. Second, it is {\em mechanical} and opens up opportunities
for automation: approximation errors for a whole program and a set
of disparate approximate transformations 
are 
generated 
%in a mechanical way, 
simply by composing the errors of
the individual transformations.  Finally, our approach
reduces the correctness of approximately transformed programs to the
much easier question of whether a generated error expression is less
than or equal to a given error bound.

The key technical insight that makes our semantics possible is that,
despite past work on using metric spaces and real numbers in program
semantics (e.g., \cite{turi98,reed87,manes86}), we argue that real
numbers cannot in general capture how, for example, the output error
of a function depends on the input and its error. Instead, our
approach allows arbitrary System F types for errors. We show that this
can accurately capture errors, for example, by using functions as
errors for functional data and polymorphic lambdas as errors for
polymorphic data. To allow this, our semantics is based around a novel
notion of an \emph{approximation type}, which is a ternary logical
relation \cite{dryer11,reynolds83} between exact expressions
$e$, approximate expressions $a$, and error expressions $q$. In
addition to the above benefits, this approach can also handle changes
of type between exact and approximate expressions, e.g., approximating
real numbers by floating-point.
% We show that our semantics can capture a wide range of past work on
% language-based approximations.

The remainder of the paper is organized as follows. Section
\ref{sec:overview} motivates our semantics with a high-level overview.
Section \ref{sec:system-f} defines our input language, \fadtsp, which
is System F with algebraic datatypes and built-in operations.  Section
\ref{sec:approximations} defines our semantics by defining the notion
of approximation types mentioned above. Section
\ref{sec:approximating-compiler} then shows how our semantics of
approximation types can be used to verify an approximating compiler
which compiles real numbers into floating-point numbers and optionally
performs loop perforation
\cite{misailovic11,sidiroglou11}. % and limit truncation.
Note that the error bounds created for the real to floating-point
approximation essentially yields a general approach to floating-point
error analysis \cite{goubault11,darulova11} that works for
higher-order and even polymorphic programs.  Finally, Section
\ref{sec:related-work} discusses related work and Section
\ref{sec:conclusion} concludes.

%%%%%%%%%%%%%%%%%%%%%%%%%%%%%%%
%%%%%       Section       %%%%%
%%%%%%%%%%%%%%%%%%%%%%%%%%%%%%%
\section{Approximating Programs}
\label{sec:overview}

The goal of this work is to give a semantics for \emph{approximate
  transformations}, which convert an exact program $e$ into an
approximate program $a$ that, although not identical to the $e$, is
within some quantifiable error bound $q$.  Our notion of approximate
transformation is very general, but it includes at least:
\begin{itemize}
\descitem{Data Approximations:} $a$ uses a less
  exact datatype, such as floating-point numbers instead of reals;
% \item[De-Noising:] $a$ performs applies a function, such as a Gaussian
%   blur, to reduce the noise in the input;
\descitem{Approximations of Incomputable Operations:}
  $a$ uses inexact but comput\-able versions of potentially
  incomputable operations, such as $f(n)$ for finite $n$ in place of
  $\lim_{x\to\infty}f(x)$; and
  %$(f(x+n)-f(x))/n$ instead of $f'(x)$; and
\descitem{Approximate Optimizations:}
  $a$ performs a cheaper, less precise version of a computation,
  such as the identity function in place of $\sin(x)$.
\end{itemize}
The point of language-based approximations is that these
transformations become part of the language semantics, which
precisely captures the relationship between exact programs,
their approximations, and the associated error bounds.

As a simple yet illustrative example, consider an approximate
transformation that replaces the $\sin$ operator with the identity
function $\lamabs{x}{x}$.
% This might be represented with the
% following rule:
% \[
% \infer{\compilestonoerr[]{\sin}{\lamabs{x}{x}}}{}
% \]
Such an approximation can greatly reduce computation, since $\sin$
(for floating-point numbers) can be a costly operation, while the
identity function is a no-op. Intuitively, we know that this change
does not greatly affect the output when $x$ is close to $0$. If,
however, the output of a call to $\sin$ is then passed to a
numerically sensitive operation, such as a reciprical, then the small
change resulting from replacing $\sin$ by the identity could lead to a
large change in the final result. Thus, our goal is to quantify the
error introduced by this approximate transformation in a local,
compositional way, allowing this error to be propogated through the
rest of the program.

\begin{figure}
  \centering
  \includegraphics[width=\linewidth]{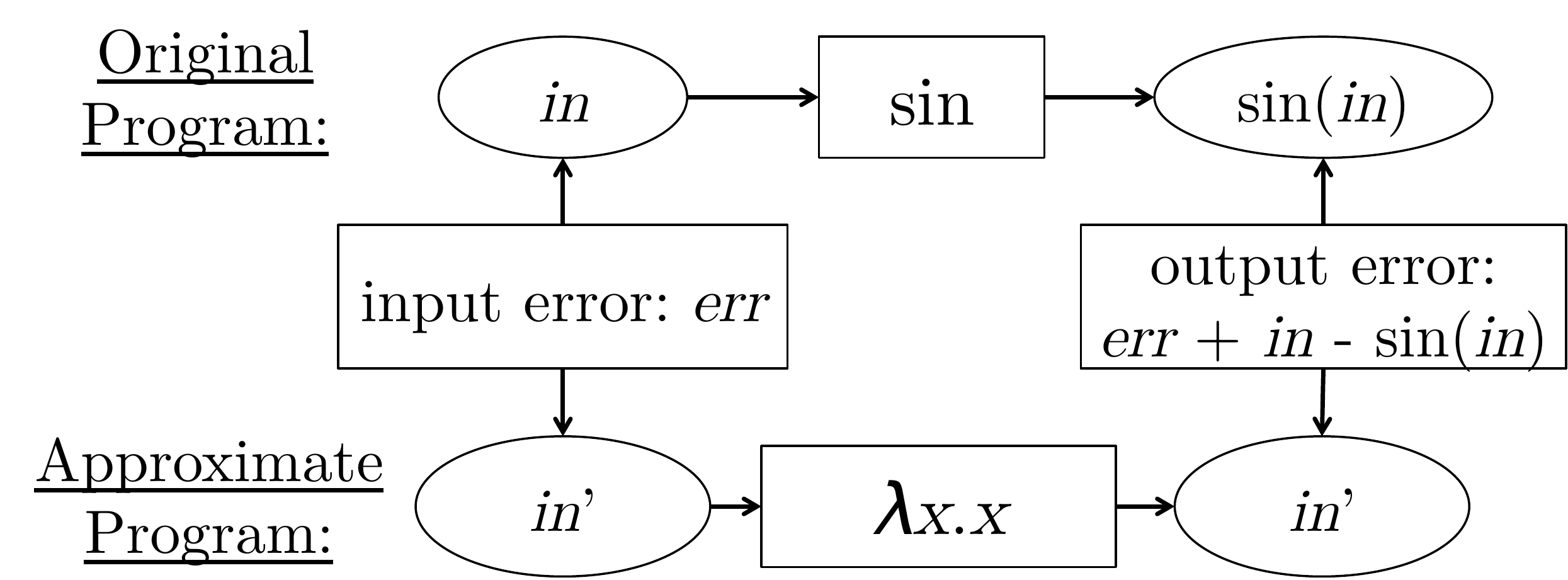}
  \caption{Error of Approximating $\sin$ with $\lamabs{x}{x}$}
  \label{fig:sin-error}
%  \vspace{-0.2in}
\end{figure}

Figure \ref{fig:sin-error} allows us to visualize the error resulting
from replacing $\sin$ with the identity function. We assume that the
input $in$ has already been approximated, yielding an approximate
input $in'$ with some approximation error $err$. The exact output,
$\sin(in)$, gets replaced by an approximate result equal to
$in'$. The error for this expression can then be calculated as
$err+in-\sin(in)$, as described in the figure.

This example leads to the following observations:
\begin{enumerate}
\item \textbf{Errors are programs:} Although the standard approach to
  quantification and errors is to use metric spaces and real numbers,
  the error expression $err+in-\sin(in)$ in our example depends on
  both the input $in$ and the input error $err$; i.e., it is a
  function, which cannot be represented by a single real number.
\item \textbf{Errors are compositional:} Approximation errors in the
  input to $\sin$ are substituted into the approximation error for
  $\sin$ itself, and this error is in turn passed to any further
  approximations that occur at a later point in the computation.
\end{enumerate}

In order to formalize the relationship between programs, their
approximations, and the resulting errors, we define a semantic notion
below called an \emph{approximation type}.  An approximation type is a
form of logical relation (see, e.g.,
\cite{dryer11,reynolds83}), satisfying certain properties
discussed below, that relates exact expressions $e$, approximate
expressions $a$, and error expressions $q$. If $\A$ is an
approximation type, we write $\appr[\A]{q}{a}$ for the set of all
exact expressions $e$ related by $\A$ to approximate expression $a$
and error expressions $q$. Thus, $e\in\appr[\A]{q}{a}$ means that
$e$ can be approximated by $a$ with approximation error no greater
than $q$ when using approximation type $\A$.

As a first example, we define the approximation type $\FloatA$ that
relates real number expressions $e$,
floating-point expressions $a$, and non-negative real number expressions
$q$ iff $q$ is the distance between the $e$ and the real number
corresponding to $a$. For instance, the real number $\pi$ is
within error $0.1415926\ldots$ of the floating-point number $3$,
meaning that
%\[
$\pi\in\appr[\FloatA]{0.1415926\ldots}{3}$
%\]
holds.
If the only difference between $a$ and $e$ is that the former uses
floating-point numbers in place of reals, then showing
$e\in\appr[\FloatA]{q}{a}$ is essentially a form of floating-point
error analysis. Using $\FloatA$ is more general, however, because it
allows the possibility that $a$ performs further approximations, e.g.,
using the identity in place of $\sin$. Taking this comparison further,
$\FloatA$ is specifically like an interval-based error analysis; we
examine this more closely in Section \ref{subsec:floating-point}.

For functions, following Figure \ref{fig:sin-error}, we use an
approximation type where the error of approximating a function is
itself a function, that maps exact inputs and their approximation
errors to output approximation errors. More formally, if $\A_1$ and
$\A_2$ are approximation types for the input and output of a function,
respectively, then then functional approximation type
$\A_1\apprto\A_2$ is defined such that
$e\in\appr[\A_1\apprto\A_2]{q}{a}$ iff
%\[
$e\;e_1\in\appr[\A_2]{q\;e_1\;q_1}{a\;a_1}$
%\]
for all $e_1$, $a_1$, and $q_1$ such that
$e_1\in\appr[\A_1]{q_1}{a_1}$.
% Note the similarity to the standard logical relation for functions.
% FIXME: mention the closed-ball explanation...?

These notions can then be used to prove correctness of approximate
tranformations, as follows. First, the designer of an approximate
transformation gives an \emph{approximation rule} for the judgment
$\compilesto[]{e}{a}{q}{\A}$, stating that expressions matching $e$
can be transformed into $a$ with error at most $q$ using approximation
type $\A$. Approximation rules are explained in more detail below,
but, as an example, our $\sin$ approximation above can be captured as:
\[
\infer{\compilesto[]{\sin}{\lamabs{x}{x}}{\lamabs{x}{\lamabs{q}{q+x-\sin(x)}}}{\FloatA\apprto\FloatA}}{}
\]
After formulating this rule, it must then be proved correct, by
proving that $e\in\appr[\A]{q}{a}$ for all $e$, $a$, and $q$ that
match the rule. As argued above, a significant benefit of this
approach is that it allows approximate transformations to be analyzed
and proved correct on their own, without reference to the programs in
which they are being used.

% FIXME: add more benefits here...?

% \begin{definition}[Approximate Correctness]
%   Expression $a$ is an \emph{approximately correct} implementation of
%   $e$ with error $q$ under approximation $\A$ iff
%   $e\in\appr[\A]{q}{a}$.
% \end{definition}

%%%%%%%%%%%%%%%%%%%%%%%%%%%%%%%
%%%%%       Section       %%%%%
%%%%%%%%%%%%%%%%%%%%%%%%%%%%%%%
\section{System F as a Logical Language}
\label{sec:system-f}

In the remainder of this paper, we work in System F with algebraic
datatypes (ADTs) (see, e.g., Pierce \cite{pierce02}), extended to
allow uncountability and incomputability. Specifically, ADTs can have
uncountably many constructors, while built-in operations can perform
potentially incomputable functions. The former is useful for modeling
the real numbers, while the latter is useful for modeling operations
on real numbers such as comparison that are incomputable in general
(e.g., \cite{edalat05,boehm86}). We refer to this language as \fadtsp.

The only additional technical machinery needed for these extensions to
System F is to require that all built-in functions are definable in
the meta-language (set theory or type theory). More specifically, we
assume as given some set of built-in operation symbols $f$, each with
a given type $\tau_f$ and meta-language relation $R_f$ relating
allowed inputs for the function defined by $f$ to their corresponding
outputs. Further, we assume that each $R_f$ obeys $\tau_f$ and relates
only one output to any given set of inputs. The small-step evaluation
relation of System F is then be extended to allow $f$ applied to any
values to evaluate to any output related to those inputs by $R_f$.
The details are straightforward but tedious, and so are omitted here.

% When using \fadtsp, we use specific letters or symbols, possibly with
% subscripts, superscripts, or primes, to denote specific sorts of
% objects; e.g., both $x$ and $x_1'$ refer to variables.  In addition, a
% super-script arrow denotes a sequence; e.g., $\seq{x}$ denotes
% $x_1,\ldots,x_n$. We use $\seqlen{x}$ to denote the length of
% $\seq{x}$.  Repeated notation is occasionally combined with other
% notation; for example, $\vec{\tau}\to\tau$ denotes the type
% $\tau_1\to\ldots\to\tau_n\to\tau$.

We use $\rrto$ to denote the small-step evaluation relation of
\fadtsp; i.e., $e\rrto e'$ means that $e$ evaluates in one step to
$e'$. Typing contexts $\Gamma$ are lists of type variables $X$ that
are considered in scope, along with pairs $x:\tau$ of expression
variables $x$ along with their types.  Substitutions for expressions
$e_1$ through $e_n$ for variables $x_1$ through $x_n$ are written
$[e_1/x_1,\ldots,e_n/x_n]$. These are represented with the letter
$\sigma$, and we define capture-avoiding substitution $\sigma e$ for
expressions and $\sigma\tau$ for types in the usual manner.
Well-formedness of types and expressions is given respectively by the
kinding judgment $\kindj{\tau}$ and typing judgment $\typej{e}{\tau}$,
defined in the standard manner. We write $\expset{\tau}$ for the set
of all expressions $e$ such that $\typej{e}{\tau}$.  Expressions or
types with no free variables are \emph{ground}.  Typing is extended to
typing contexts $\Gamma$ and substitutions in the usual manner:
$\wfj[]{\Gamma}$ indicates that all types in $\Gamma$ are well-kinded,
while $\typej{\sigma}{\Gamma'}$ indicates that
$\Dom(\sigma)=\Dom(\Gamma')$, $\kindj{\sigma(X)}$ for all
$X\in\Dom(\sigma)$, and $\typej{\sigma(x)}{\Gamma'(x)}$ for all
$x\in\Dom(\sigma)$.  We assume all types are well-kinded and all
expressions, contexts, and substitutions are well-typed below.

We write $\term{e}$ to denote that $e$ is terminating.  An
\emph{expression context} $C$ an expression with exactly one
occurrence of a ``hole'' \tctxemp, and $\tctx{C}{e}$ denotes the
(non-capture-avoiding) replacement of $\tctxemp$ with $e$.  Contextual
equivalence $e_1\ctxeq e_2$ is then defined to hold iff, for all $C$
and $\tau'$ such that $\typej[\cdot]{\tctx{C}{e_i}}{\tau'}$ for
$i\in\{1,2\}$, we have that $\term{\tctx{C}{e_1}}$ iff
$\term{\tctx{C}{e_2}}$.
We use $\bot$ to denote an arbitrary non-terminating expression.
$\fixappnot{\lamabs{x}{x}}$ of a given type $\tau$.
% We assume the following ADTs: the
% unit type $\Unit$ with the sole constructor $\unit::\Unit$;
% %the Boolean type $\Bool$ with constructors $\true$ and $\false$;
% % the product type $\tau_1\times\tau_2$ with constructor
% % $\tuple{}::\foralltp{X}{\foralltp{Y}{X\to Y\to X\times Y}}$, where we
% % often write $\pair{e_1}{e_2}$ for the application of this constructor
% % to arguments $e_1$ and $e_2$, omitting the type arguments;
% the list type $\List\;\tau$, with constructors
% $\nil::\foralltp{X}{\List\;X}$ and
% $\consname::\foralltp{X}{X\to\List\;X\to\List\;X}$, where we write
% $\mklist{e_1,\ldots,e_n}$ for the list
% with elements $e_1$ through $e_n$;
% % We write $\head$ and $\tail$ for
% % the usual list functions, defined using \mycasename; these return $\bot$
% % when applied to the empty list \nil.
% the type \Fixed\ of fixed-point numbers with some given precision; the
We assume below
type \R\ of real numbers, represented with an uncountable number of
constructors, and the type \Rpinf\ of the non-negative reals with
infinity.  We use $\zeroR$ for the \R-constructor corresponding to the
real number 0, and $\leqR$, $\plusR$, and $\dR$ for the function
symbols whose functions perform real number comparison, addition, and
absolute difference, both on \R\ and, abusing notation slightly, on
\Rpinf.

Since \fadtsp\ can contain incomputable built-in operations, we can
additionally use it as a meta-logic by embedding any relations of a
given meta-language, such as set theory or type theory, as built-in
operations. We assume an ADT $\Prop$ with the sole constructor
$\top::\Prop$, which will intuitively be used as the type of
propositions; a ``true'' proposition terminates to $\top$ and a
``false'' one is non-terminating, i.e., is contextually equivalent to
$\bot$. A meta-language relation $R$ can then be added to \fadtsp\ as
a function symbol $f_R$ of type
$\foralltp{\vec{X}}{\vec{\tau}\to\Prop}$ iff $R$ is a set of tuples
$\pair{\seq{\tau'}}{\seq{e}}$ such that
$\typej[\cdot]{\tpapp{f_R}{\seq{\tau'}}\;\seq{e}}{\Prop}$ that is
closed under contextual equivalence and contains only terminating
expressions $\seq{e}$.  Note that this latter restriction is not too
significant because we can always use a ``thunkified'' relation $R'$
of type
$\foralltp{\vec{X}}{(\Unit\to\tau_1)\to\ldots\to(\Unit\to\tau_n)\to\Prop}$
such that $\pair{\vec{\tau'}}{\vec{e}}\in R$ iff
$\pair{\vec{\tau'}}{\lamabs{x}{e_1},\ldots,\lamabs{x}{e_n}}\in R$.  As
an example, we can immediately see that the thunkified version of
contextual equivalence itself is a relation of type
$\foralltp{X}{(\Unit\to X)\to(\Unit\to X)\to\Prop}$.

\begin{lemma}[Consistency]
  For any class of built-in operations $f$ with functions $F_f$,
  $\top\ctxeq\bot$ does not hold.
\end{lemma}

In the below, $\phi$ refers to expressions of type
$\foralltp{\vec{X}}{\vec{\tau}\to\Prop}$.  A \emph{constrained
  context} is a pair $\cctxdef$ of a context $\Gamma$ and an
expression $\phi$ such that $\typej{\phi}{\Prop}$.  A substitution
$\sigma$ \emph{satisfies} $\cctxdef$, written
$\ssatj{\sigma}{\cctxdef}$, iff $\typej[\cdot]{\sigma}{\Gamma}$ and
$\sigma(\phi)\ctxeq\top$. We say that $\cctxdef$ \emph{entails}
$\phi$, written $\entj{\phi}$, iff
$\forall\ssatj{\sigma}{\cctxdef}.\sigma(\phi)\ctxeq\top$ holds.
Finally, we say that $\sigma$ is a substitution \emph{from
  $\cctxdef[1]$ to $\cctxdef[2]$}, written
$\typejcctxdef{2}{\sigma}{1}$, iff
$\typej[\Gamma_2]{\sigma}{\Gamma_1}$ and
$\entjprot{\cctxdef[2]}{\sigma(\phi_1)}$.

% \begin{lemma}
%   If $\typejcctxdef{2}{\sigma_1}{1}$ and $\typejcctxdef{3}{\sigma_2}{2}$,
%   then $\typejcctxdef{3}{\sigma_2\circ\sigma_1}{1}$.
% \end{lemma}

% FIXME: other elucidating lemmas?

%%%%%%%%%%%%%%%%%%%%%%%%%%%%%%%
%%%%%       Section       %%%%%
%%%%%%%%%%%%%%%%%%%%%%%%%%%%%%%
\section{A Semantics of Program Approximation}
\label{sec:approximations}

In this section, we define approximation types and illustrate them
with examples. The main technical difficulty is that the
straightforward way to handle free (expression and type) variables
involves a circular definition.  Specifically, to fit with the logical
relations approach, $e\in\appr[\A]{q}{a}$ for $e$, $a$, and $q$ with
free variables should only hold iff %
$\sigma e\in\appr[\A]{\sigma q}{\sigma a}$ for all substitutions
$\sigma$ for exact variables $\xe$, approximate variables $\xa$, and
error variables $\xq$ such that
$\sigma\xe\in\appr[\A']{\sigma\xq}{\sigma\xa}$ for some approximation
$\A'$. This defines approximation types in terms of approximation
types!  To remove this circularity, we first define approximation
types that handle a given set of free variables arbitrarily, in
Section \ref{subsec:approximation-types}. We then show in Section
\ref{subsec:approximation-families} how to lift ground approximation
types into approximation functions, which uniformly handle any given
context of variables in the ``right'' way.

%%%%%%%%%%%%%%%
% Sub-section %
%%%%%%%%%%%%%%%
\subsection{Approximation Types}
\label{subsec:approximation-types}

\begin{definition}[Expression-Preorder]
  Let $\kindj{\tau}$. A relation $\leq$ is called a
  \emph{$\ctxtp{\Gamma}{\tau}$-expression preorder} iff it is a preorder
  (reflexive and transitive) over
  $\ctxeq$-equivalence classes of $\expset{\tau}$.
  % Stated differently,
  % the following must hold:
  % \begin{itemize}
  % \descitem{Typing:} If $e_1\leq e_2$ then $\typej{e_1}{\tau}$ and $\typej{e_2}{\tau}$;
  % \descitem{Reflexivity:} If $e_1\ctxeq e_2$ for $e_1,e_2\in\expset{\tau}$ then $e_1\leq e_2$; and
  % \descitem{Transitivity:} $e_1\leq e_2$ and $e_2\leq e_3$ implies $e_1\leq e_3$.
  % %\descitem{Substitution:} $e_1\leq e_2$ implies $\sigma e_1\leq\sigma e_2$ for all well-typed $\sigma$.
  % \end{itemize}
\end{definition}

\begin{definition}[Quantification Type]
  \label{def:quantification-type}
  A Quantification type is a is a tuple
  $\Q=\quanttctx{\Gamma}{Q}{\leq}{+}{0}$ of: a typing context $\Gamma$
  and a type $Q$ such that $\kindj{Q}$; a
  $\ctxtp{\Gamma}{Q}$-expression preorder $\leq$; and two expressions
  $+$ (sometimes written in infix notation) and $0$ of types $Q\to
  Q\to Q$ and $Q$ (relative to $\Gamma$), respectively, such that $0$
  is a least element for $\leq$, $+$ is monotone with respect to
  $\leq$ and $\triple{\expset{Q}}{+}{0}$ forms a monoid with respect
  to the equivalence relation $\lesseqgtr=(\leq\cap\geq)$. Stated
  differently, the following must hold:
  \begin{itemize}
  % \descitem{Reflexivity:} $e\ctxeq e'$ implies $e\lesseqgtr e'$;
  % \descitem{Transitivity:} $e_1\leq e_2\leq e_3$ implies $e_1\leq e_3$;
  \descitem{Closedness:} $e_1,e_2\in\expset{Q}$ implies
    $e_1+e_2\in\expset{Q}$;
  \descitem{Monotonicity:} $e_1\leq e_1'$ and $e_2\leq e_2'$ implies
    $e_1+e_2 \leq e_1'+e_2'$;

  \descitem{Leastness of $0$:} $0\leq e$ for all $e\in\expset{Q}$;
  %\descitem{Boundedness:} $0\leq e\leq\infty$ for all $e\in\field{\leq}$;
  \descitem{Identity:} $e+0 \lesseqgtr e$ for all $e\in\expset{Q}$;
  %\descitem{Per-Element Identity:} $\forall e\in\field{\leq}.e\lesseqgtr e+(I\;e)$;

  %\descitem{Absorbtion:} $e+\bot \lesseqgtr \bot$ for all $e\in\expset{Q}$;

  \descitem{Commutativity:} $e_1+e_2\lesseqgtr e_2+e_1$ for $e_i\in\expset{Q}$; and
  \descitem{Associativity:} $e_1+(e_2+e_3)\lesseqgtr(e_1+e_2)+e_3$ for $e_i\in\expset{Q}$.
%  \descitem{Descending Chain:} $e_1,e_2\in\field{\leq}$ implies $e_1+(\downQ[]{e_1}{e_2})\lesseqgtr e_1$ or $e_1+(\downQ[]{e_1}{e_2})< e_1+e_2$.

  \end{itemize}
  Such a tuple is sometimes called a $Q$- or
  $\ctxtp{\Gamma}{Q}$-Quantification type.  If $\Gamma=\cdot$ then \Q\
  is said to be \emph{ground}.
\end{definition}

In the below, we often omit $\Gamma$ when it is clear from context,
writing $\quantt{Q}{\leq}{+}{0}$.  We also write $\QQ$, $\leqQ$,
$\plusQ$, and $\IQ$ for the corresponding elements of $\Q$. We
sometimes omit the $\Q$ subscript where $\Q$ can be inferred from
context.
% Abusing notation slightly, we also sometimes write $\Q$ for the type
% component $\QQ$ of \Q.
The following lemma gives a final, implied constraint,
that $\bot$ is always infinity:

\begin{lemma}
  \label{lemma:quant-types}
  For any quantification type $\Q$, $e\leqQ\bot$ for all $e\in\expset{Q}$.
\end{lemma}

\begin{example}[Non-Negative Reals]
  $\QR=\quantt{\Rpinf}{\leqR}{\plusR}{\zeroR}$ is a
  ground quantification type that corresponds to the standard
  ordering and addition on the non-negative reals.
\end{example}

% \begin{example}[Lists of Non-Negative Reals]
%   $\QR=\quantt{\Rpinf}{\leq_{\List\;R}}{\mathsf{zipWith}\;\plusR}{\zeroR}$ is % %   ground quantification type.
% \end{example}

\begin{example}[Non-Negative Real Functions]
  \label{ex:q-rfun}
  For any ground $\tau$,
  %\[
  $\quantt{\tau\to\Rpinf}{\leq_{\tau\to\Rpinf}}{\lamabs{x_1}{\lamabs{x_2}{\lamabs{y}{x_1\;y\plusR x_2\;y}}}}{\lamabs{y}{\zeroR}}$
  %\]
  is a ground quantification type, where addition is performed pointwise
  on functions and $e_1\leq_{\tau\to\Rpinf}e_2$ iff
  $e_1\;e\leqR e_2\;e$ for all ground $e$ of type $\tau$; i.e.,
  iff the output of $e_1$ is always no greater than that of $e_2$.
  % The straightforward generalization to multiple types $\vec{\tau}$
  % is called $\Q_{\vec{\tau}\to\Rpinf}$.
\end{example}

% %%%%%%%%%%%%%%%
% % Sub-section %
% %%%%%%%%%%%%%%%
% \subsection{Approximate Equality}
% \label{subsec:approx-eq}

% For a given quantification type \Q,
% % which intuitively defines a notion of distance,
% we define the notion of approximate equality relative to \Q\ as
% follows:

\begin{definition}[Approximate Equality]
  \label{def:approximate-equality}
  Let $\Gammaq$ and $\Gammae$ be any contexts with disjoint domains,
  let \Q\ be a $\ctxtp{\Gammaq,\Gammae}{Q})$-quantification type, and let $E$
  by any type such that $\kindj[\Gammae]{E}$.  A
  \emph{$(\Q,\Gammae,E)$-approximate equality relation} is a ternary
  relation
  $\aeq{q}{e_1}{e_2}\subseteq\expsetqe{Q}\times\expsete{E}\times\expsete{E}$,
  where $q\in\expsetqe{Q}$ and each $e_i\in\expsete{E}$, that satisfies
  the following:
  \begin{itemize}
  \descitem{Upward Closedness:} $\aeq{q}{e_1}{e_2}$ and $q\leq q'$ implies
    $\aeq{q'}{e_1}{e_2}$;

  % \descitem{Substitution:} $\aeq{q}{e_1}{e_2}$ implies
  %   $\aeq{\sigma q}{\sigma e_1}{\sigma e_2}$ for all $\sigma$;

  %\descitem{Completeness:} $\aeq{\inftyQ}{e_1}{e_2}$ for all $e_1,e_2\in S$;

  \descitem{Reflexivity:} $e_1\ctxeq e_2$ implies $\aeq{\zeroQ}{e_1}{e_2}$;
  %\descitem{Reflexivity:} $e_1\ctxeq e_2$ implies $\aeq{q}{e_1}{e_2}$ for all $q\in\expset{Q}$;

  \descitem{Symmetry:} $\aeq{q}{e_1}{e_2}$ implies $\aeq{q}{e_2}{e_1}$;

  \descitem{Triangle Inequality:} $\aeq{q_1}{e_1}{e_2}$ and $\aeq{q_2}{e_2}{e_3}$
    implies $\aeq{q_1+q_2}{e_1}{e_3}$; and

  \descitem{Completeness:} $\aeq{\bot}{e_1}{e_2}$ for all $e_1,e_2\in\expset{E}$.
%   \descitem{Bottom Correctness:} $\aeq{q}{e}{\bot}$ iff $e\ctxeq\bot$ or $q\ctxeq\bot$.

%   \descitem{Limit Closedness:} If $q_1>q_2>\ldots$ is an infinite descending
%     chain with greatest lower bound $q$, and if $\forall i.\aeq{q_i}{e_1}{e_2}$,
%     then $\aeq{q}{e_1}{e_2}$.

  \end{itemize}
\end{definition}

An approximation equality relation is \emph{ground} iff
$\Gammae=\Gammaq=\cdot$.
% Upward closedness means that
% that approximate equality still holds if the error bound $q$ is
% weakened.  Reflexivity states that the distance between equivalent
% expressions is $\zeroQ$.  Symmetry ensures that distance is not
% directional.  The Triangle Inequality essentially gives meaning to $+$
% by stating that $+$ can be used to bound the distance between $e_1$
% and $e_3$ when the distance from both of these to $e_2$ is known.
% Finally, Completeness says that all
% expressions are related at least by the infinite error $\bot$.
In the below, we often omit $\Gammaq$ and $\Gammae$ when clear from
context.  We use a subscript to denote which approximate equality
relation is intended, as in $\aeq[\A]{q}{e_1}{e_2}$, when it is not
clear from context.

\begin{example}[Reals]
  The relation $\aeq[\R]{q}{e_1}{e_2}$ that holds iff $q\ctxeq\bot$,
  or $e_1\ctxeq e_2$, or $|e_1-e_2|\leqR q$ is a ground
  $(\Rpinf,\R)$-approximate equality relation that corresponds to
  the standard distance over the reals.
\end{example}

% \begin{example}[Lists of Reals]
%   The relation $\aeq[\LoneA{\R}]{q}{e_1\!}{\!e_2}$ that holds iff
%   $q\ctxeq\bot$, or $e_1\ctxeq e_2$, or each
%   $e_i\ctxeq\mklist{e_{i,1},\ldots,e_{i,n}}$ such that
%   $(\sum_{j=1}^{n}|e_{1,j}-e_{2,j}|)\leqR q$, is a ground
%   $(\Rpinf,\List\;\R)$-approximate equality relation.
% \end{example}

\begin{example}[Real Functions]
  \label{ex:real-functions}
  The relation $\aeq[\R\to\R]{q}{e_1}{e_2}$ that holds iff
  $q\ctxeq\bot$, or $e_1\ctxeq e_2$, or $|(e_1\;r)-(e_2\;r)|\leqR
  (q\;r\;r^+)$ for all $r\in\R$ and $r^+\in\Rpinf$, is a ground
  $(\R\to\Rpinf\to\Rpinf,\R\to\R)$-approximate equality relation,
  where, intuitively, the distance between two functions $e_1$ and
  $e_2$ is given by a function $q$ that bounds the distance between
  their outputs for each input.
\end{example}

% %%%%%%%%%%%%%%%
% % Sub-section %
% %%%%%%%%%%%%%%%
% \subsection{Approximations}
% \label{subsec:approx}

% We now define the precise relationship between exact and approximate
% expressions as an \emph{approximation}:

% At a high level, approximations capture the relationship between, for
% example, the floating-point numbers and the rationals, where each
% floating-point \emph{approximates} the set of rationals closest to
% it. Intuitively, an approximation is a relationship between an
% \emph{exact set} of values being approximated and an \emph{approximate
%   set} of values that are used to approximate subsets of the exact
% set. In our example, the rational numbers are the exact set, while the
% floating-point numbers are the approximate set. We use \emph{exact
%   element} and \emph{approximate element} for members of these two
% sets. More formally, we define approximations as follows:

\begin{definition}[Approximation Type]
  \label{def:approximation}
  Let $\Gammae$, $\Gammaa$, and $\Gammaq$ be three domain-disjoint
  contexts.  A \emph{$(\Gammae,\Gammaa,\Gammaq)$-approximation type}
  is a tuple $\apprtuple{\Q}{E}{A}{\aeqrel}{\appr{\cdot}{\cdot}}$ of:
  a $\ctxtp{\Gammaq,\Gammae}{Q}$-quantification type \Q; types $E$ and
  $A$, called respectively the exact and approximate types, such that
  $\kindj[\Gammae]{E}$ and $\kindj[\Gammaa]{A}$; a
  $(\Q,\Gammae,E)$-approximate equality relation $\aeqrel$; and a
  mapping
  $\appr{q}{a}:(\expsetqe{Q}\times\expseta{A})\to\powerset(\expsete{E})$,
  where $q\in\expsetqe{Q}$ and $a\in\expseta{A}$, that satisfies:
  \begin{itemize}
  \descitem{Error Weakening:} $q_1\leq q_2$ implies $\appr{q_1}{a}\subseteq\appr{q_2}{a}$;

  \descitem{Error Addition:} $e_1\in\appr{q}{a}$ and
    $\aeq{q'}{e_1}{e_2}$ implies $e_2\in\appr{q+q'}{a}$\!;

  % \descitem{Substitution:} $e\in\appr{q}{a}$ implies $\sigma e\in\appr{\sigma q}{\sigma a}$;

  \descitem{Equivalence:} If $q\ctxeq q'$, $a\ctxeq a'$, and $e\ctxeq e'$ then $e\in\appr{q}{a}$ implies $e'\in\appr{q'}{a'}$;

  \descitem{Approximate Equality:} $e_1,e_2\in\appr{q}{a}$ implies $\aeq{q+q}{e_1}{e_2}$; and

  % \descitem{Completeness:} $e\in\appr{\bot}{a}$ for all $e\!\in\!\expsete{E}$ and
  %   $a\!\in\!\expseta{A}$.
%   \descitem{Bottom Correctness:} $\bot\in\appr{a}{q}$ iff $a\ctxeq\bot$ or $q\ctxeq\bot$.

%   \descitem{Limit Closedness:} If $q_1>q_2>\ldots$ is an infinite
%     descending chain with greatest lower bound $q$, then
%     $\appr{q}{a}=\cap_i\appr{q_i}{a}$.

  \end{itemize}
\end{definition}

A \emph{ground} approximation type is one where
$\Gammae=\Gammaa=\Gammaq=\cdot$.
% Error Weakening means that greater errors lead to super-sets. Error
% Addition states that adding some $q'$ to the error adds all exact
% elements within distance $q'$.  Equivalence requires approximate
% sets to be closed under $\ctxeq$ and for equal approximate elements
% $a$ to define the same sets. Note that Error Weakening already
% handles $\ctxeq$ on $q$, that is, it implies that
% $\appr{q}{a}=\appr{q'}{a}$ whenever $q\ctxeq q'$. Approximate
% Equality ensures that the diameter of the set $\appr{q}{a}$, or
% maximum distance between elements, is $q+q$.  Note that, by
% Completness (for Approximate Equalities), Error Addition, and Lemma
% \ref{lemma:quant-types}, $\appr{\bot}{a}=\expset{E}$.
% % Finally, Bottom
% % Correctness makes the following stipulations: the approximate sets for
% % $\bot$ with non-infinite error to contain only $\bot$; all other
% % approximate sets with non-infinite error to not contain bottom; and
% % all infinite error approximate sets $\appr{\bot}{a}$ to contain all
% % exact expressions $e$.
% % % Finally, Limit Closedness states, at a high level, that
% % % $\appr{q}{a}$ equals $\lim_{x\to q}\appr{x}{a}$.
In the below, we write \A\ for approximation types, writing
$\appr[\A]{q}{a}$ for the set to which \A\ maps $q$ and $a$,
$\aeq[\A]{\cdot}{\cdot}{\cdot}$ for the approximate equality relation,
$\EA$ and $\AA$ for the exact and approximate types,
$\QA$ for the quantification type, and $\QQA$, $\leqA$, and $\plusA$
for the elements of $\QA$. Again, we often omit the subscript \A\ when
it can be inferred from context.

\begin{example}[Floating-Point Numbers]
  \label{example:floating-point}
  Let $e\in\appr[\FloatA]{q}{a}$ iff $q\ctxeq\bot$, or $e\ctxeq\bot$
  and $a\ctxeq\bot$, or $|e-\mathsf{real}(a)|\leqR q$ where
  $\mathsf{real}$ maps (the value of) $a$ to its corresponding real
  number. We then have that
  $\FloatA=\apprtuple{\QR}{\R}{\FloatA}{\aeqrel[\R]}{\appr[\FloatA]{\cdot}{\cdot}}$
  is a ground approximation type where a real can be approximated by a
  floating-point number with an error given by the real-number
  distance between the two.
\end{example}

% \begin{example}[Lists of Fixed-Point Numbers]
%   Let $e\in\appr[\LoneA{\FloatA}]{q}{a}$ iff $q\ctxeq\bot$, or $e\ctxeq\bot$
%   and $a\ctxeq\bot$, or $e\ctxeq\mklist{e_1,\ldots,e_n}$ and
%   $a\ctxeq\mklist{a_1,\ldots,a_n}$ such that
%   $(\sum_{i=1}^{n}|a_{i}-e_{i}|)\leqR q$, viewing (the value of) each $a_i$
%   as its corresponding real number. We then have that
%   $\LoneA{\FloatA}=\apprtuple{\QR}{\List\;\R}{\List\;\FloatA}{\aeqrel[\LoneA{\R}]}{\appr[\LoneA{\FloatA}]{\cdot}{\cdot}}$
%   is a ground approximation.
% \end{example}

\begin{example}[Floating-Point Functions]
  \label{example:floating-point-functions}
  Let $e\in\appr[\FloatA\apprto\FloatA]{q}{a}$ iff $q\ctxeq\bot$, or
  $e\!\ctxeq\!\bot$ and $a\!\ctxeq\!\bot$, or
  $e\,e'\in\appr[\FloatA]{q\,e'\,q'}{a\,a'}$ for all
  $e'\in\appr[\FloatA]{q'}{a'}$. We then have that
  $\FloatA\apprto\FloatA=
  %\[
  \apprtuple{\R\apprto\Rpinf\apprto\QR}{\R\to\R}{\FloatA\to\FloatA}{\aeqrel[\FloatA\apprto\FloatA]}{\appr[\FloatA\apprto\FloatA]{\cdot}{\cdot}}
  %\]
  $ %
  is a ground approximation type, where $\R\apprto\Rpinf\apprto\QR$ is
  the quantification type obtained from $\QR$ by applying the
  construction of Example \ref{ex:q-rfun} twice. Intuitively, this
  approximation type allows a real function $f$ to be approximated
  by a floating-point function $f'$ with an error $q$ whenever
  $q\;r\;q_r$ bounds the error between calling $f$ on exact real
  number $r$ and calling $f'$ on a floating-point number with
  at most distance $q_r$ from $r$.
\end{example}

% \begin{example}[Exact Approximation]
% \end{example}

% \begin{example}[Floating-Point Numbers]
%   For any floating-point number $f$, let $S_f$ be the set containing
%   just the real value of $f$. We then have that $\SetA{\aeqrel}{S}$ is
%   a ground approximation, where $\appr{q}{f}$ is the set of all real
%   numbers within distance $q$ of (the real number represented by) $f$.
% \end{example}

%%%%%%%%%%%%%%%
% Sub-section %
%%%%%%%%%%%%%%%
\subsection{Approximation Families}
\label{subsec:approximation-families}

% Approximation families are classes of approximations that can be
% relativized to different contexts $\Gamma$, which in turn allow us to
% define notions like Weakening and Substitution for approximation types:

\begin{definition}[Approximation Families]
  Let $\Gamma=\Gammae,\Gammaa,\Gammaq,\GammaA$ for four
  domain-disjoint typing contexts and $\phi$ be an expression such
  that $\typej{\phi}{\Prop}$. We say that
  $\F=\sextuple{E}{A}{Q}{0}{+}{F}$ is a
  \emph{$(\Gammae,\Gammaa,\Gammaq,\GammaA,\phi)$-approximation family}
  iff $\kindj[\Gammae]{E}$, $\kindj[\Gammaa]{A}$,
  $\kindj[\Gammae,\Gammaq]{Q}$, $\typej[\Gammae,\Gammaq]{0}{Q}$,
  $\typej[\Gammae,\Gammaq]{0}{Q\to Q\to Q}$, and $F$ is a
  meta-language function from substitutions $\sigma$ such that
  $\ssatj{\sigma}{\cctxdef}$ to ground approximation types $\A$ such
  that
  $\EA\!=\!\sigma(E)$, $\AA\!=\!\sigma(A)$, $\QQA\!=\!\sigma(Q)$,
  $\zeroA\!=\!\sigma(0)$, and $\plusA\!=\!\sigma(+)$.
\end{definition}

We use a subscript $\F$ to denote the elements of $\F$; e.g., $\EF$
denotes the exact type $E$ of $\F$. We write $\F(\sigma)$ for the
approximation resulting from applying the $F$ component to $\sigma$. A
$(\cdot,\cdot,\cdot,\cdot,\truepred)$-approximation family is called
\emph{ground}. Note that the ground $\F$ are isomorphic to the ground
approximation types $\A$, since, if $\F$ is ground, then the domain of
$F_\F$ consists of the sole pair $\pairsemi{\cdot}{\cdot}$.

We define \emph{approximation contexts} $\apprctx$ with grammar:
\[
%\begin{array}{l}
  %\apprctx ::=
  \cdot \bor \apprctx,\aelemdef \bor \apprctx,\atelemdef \bor \apprctx,\phielemdef
%\end{array}
\]
The form $\aelemdef$ introduces variables $\xe$, $\xa$, and $\xq$ such
that $\xa$ is an approximation of $\xe$ with error $\xq$ in some
approximation returned by approximation family \F.  The form
$\atelemdef$ introduces type variables $\Xe$, $\Xa$, and $\Xq$, along
with a variable $\xi$ that quantifies over approximations of $\Xe$ by
$\Xa$ with error $\Xq$.  Finally, the form $\phielemdef$ introduces
additional variables in $\Gamma$ and constraint $\phi$.

More formally, let $\ectx$, $\actx$, $\qctx$, and $\Actx$ be typing
contexts that contain, respectively: all $\xe$ and $\Xe$; all $\xa$
and $\Xa$; all $\xq$, $\Xq$, and contexts $\Gamma$ in a $\phielemdef$
form; and all $\xi$ variables in $\apprctx$. Further, let $\phictx$ be
the conjunction of the following formulas: $\xe\in\appr{\xq}{\xa}$ for
each $\aelemdef$; the formula $\tpapp{\isapprox}{\Xe,\Xa,\Xq}$ stating
that $\xi$ is an approximation of $\Xe$ by $\Xa$ with error $\Xq$ for
each $\atelemdef$; and $\phi$ for each $\phielemdef$. We use the
abbreviations $\eqctx=\ectx,\qctx$ and
$\eaqActx=\ectx,\actx,\qctx,\Actx$.
% FIXME: define notation \F(\apprctx)
The approximation context $\apprctx$ is well-formed, written
$\wfappr{\apprctx}$, iff $\wfj[]{\eaqActx}$ and
$\typej[\eaqActx]{\phi}{\Prop}$.
% The approximation context $\apprctx$ is well-formed, written
% $\wfappr{\apprctx}$, iff for each prefix $\apprctx_1,\aelemdef$ of
% $\apprctx$ we have that $\typejF[\apprctx_1]{\F}$, and for each
% prefix $\apprctx_2,\phielemdef$ we have that
% $\typejpred[\apprctx_2]{\phi}$.
We say $\F$ is a \emph{$\apprctx$-approximation family}, written
$\typejF{\F}$, iff $\F$ is a
$(\ectx,\actx,\qctx,\Actx,\phictx)$-approximation family.  A $\sigma$
is a \emph{substitution from $\apprctx$ to $\apprctx'$}, written
$\typej[\apprctx']{\sigma}{\apprctx}$, iff
$\typej[\cctxapprprot{\apprctx'}]{\sigma}{\cctxapprprot{\apprctx}}$.

Although the functions $F$ in approximation families
return only ground approximations, we can create non-ground
approximation types from $\F$ as follows.
Let $\apprctx$ be any approximation context and $\F$ be
any $\apprctx$-approximation family. The notation $\CtxA{\F}$
then denotes the $(\ectx,\actx,\qctx)$-approximation
$\apprtuplesm{\Q}{E}{A}{\aeqrel}{\appr{\cdot}{\cdot}}$ where
$\Q=\quanttctx{\eqctx}{\QQF}{\leq}{\zeroF}{\plusF}$ and:
\begin{itemize}
\item $q_1\leq q_2$ iff $\forall\ssatjapprctx{\sigma}.\;\sigma q_1\leq_{\F(\sigma)}\sigma q_2$;
\item $\aeq{q}{e_1}{e_2}$ iff $\forall\ssatjapprctx{\sigma}.\;\aeq[\F(\sigma)]{\sigma q}{\sigma e_1}{\sigma e_2}$; and
\item $e\in\appr{q}{a}$ iff
  $\forall\ssatjapprctx{\sigma}.\;\sigma e\in\appr[\F(\sigma)]{\sigma q}{\sigma a}$.
\end{itemize}
We call an approximation formed this way a $\apprctx$-approximation.
% Note that the last condition is identical to the condition required
% in the $\sigma$ typing rule for $\aelemdef$.

\begin{theorem}
  If $\wfappr{\apprctx}$ and $\typejF{\F}$ then $\CtxA{\F}$ is a
  valid approximation type.
\end{theorem}

\begin{lemma}[Approximation Weakening]
  Let $\typejF{\F}$ and $\wfappr{\apprctx,\apprctx'}$. We then have
  that: $\typejF[\apprctx,\apprctx']{\F}$; $q_1\leq_{\CtxA{\F}}q_2$
  implies $q_1\leq_{\CtxA[\apprctx,\apprctx']{\F}}q_2$;
  $\aeq[\CtxA{\F}]{q}{e_1}{e_2}$ implies
  $\aeq[\CtxAprot{\apprctx,\apprctx'}{\F}]{q}{e_1}{e_2}$; and
  $e\in\appr[\CtxAprot{\apprctx}{\F}]{q}{a}$ implies
  $e\in\appr[\CtxAprot{\apprctx,\apprctx'}{\F}]{q}{a}$.
  % \begin{enumerate}
  % \item $\typejF[\apprctx,\apprctx']{\F}$;
  % \item $q_1\leq_{\CtxA{\F}}q_2$ implies $q_1\leq_{\CtxA[\apprctx,\apprctx']{\F}}q_2$;
  % \item $\aeq[\CtxA{\F}]{q}{e_1}{e_2}$ implies
  %   $\aeq[\CtxAprot{\apprctx,\apprctx'}{\F}]{q}{e_1}{e_2}$; and
  % \item $e\in\appr[\CtxAprot{\apprctx}{\F}]{q}{a}$ implies
  %   $e\in\appr[\CtxAprot{\apprctx,\apprctx'}{\F}]{q}{a}$.
  % \end{enumerate}
\end{lemma}

We define substitution $\sigma(\F)$ into approximation
families as yielding the approximation family
$\sextuple{\sigma E}{\sigma A}{\sigma Q}{\sigma 0}{\sigma +}{\lamabs{\sigma'}{F(\sigma'\circ\sigma)}}$:

\begin{lemma}[Approximation Substitution]
  If $\typejF[\apprctx,\apprctx']{\F}$ and
  $\typej[\apprctx]{\sigma}{\apprctx'}$ then:
  $\typejF[\apprctx]{\sigma(\F)}$;
  $q_1\leq_{\CtxA[\apprctx,\apprctx']{\F}}q_2$ implies $\sigma
  q_1\leq_{\CtxA{\sigma(\F)}}\sigma q_2$;
  $\aeq[\CtxAprot{\apprctx,\apprctx'}{\F}]{q}{e_1}{e_2}$ implies
  $\aeq[\CtxA{\sigma(\F)}]{\sigma q}{\sigma e_1}{\sigma e_2}$; and
  $e\in\appr[\CtxAprot{\apprctx,\apprctx'}{\F}]{q}{a}$ implies $\sigma
  e\in\appr[\CtxAprot{\apprctx}{\sigma(\F)}]{\sigma q}{\sigma a}$.
  % \begin{enumerate}
  % \item $\typejF[\apprctx]{\sigma(\F)}$;
  % \item $q_1\leq_{\CtxA[\apprctx,\apprctx']{\F}}q_2$ implies
  %   $\sigma q_1\leq_{\CtxA{\sigma(\F)}}\sigma q_2$;
  % \item $\aeq[\CtxAprot{\apprctx,\apprctx'}{\F}]{q}{e_1}{e_2}$ implies
  %   $\aeq[\CtxA{\sigma(\F)}]{\sigma q}{\sigma e_1}{\sigma e_2}$; and
  % \item $e\in\appr[\CtxAprot{\apprctx,\apprctx'}{\F}]{q}{a}$ implies
  %   $\sigma e\in\appr[\CtxAprot{\apprctx}{\sigma(\F)}]{\sigma q}{\sigma a}$.
  % \end{enumerate}
\end{lemma}

\begin{definition}[$\Pi$-Approximations]
  \label{def:pi-approximation}
  If $\F$ is a $\apprctx',\apprctx$-approximation family, then
  $\piappr{\apprctx}{\F}$ is the $\apprctx'$-approximation family
  %\[
  $\sextuple{\ectx\!\to\!\EF}{\!\actx\!\to\!\AF}{\!\eqctx\!\to\!\QQF}{\!\lamabs{\eqctx}{\zeroF}}{\!\lamabs{\eqctx}{\!\plusF}}{\!F'}$
  %\]
  where $F'(\sigma)$ for $\ssatjapprctx[\apprctx']{\sigma}$ is defined
  such that: $q_1\leq_{F'(\sigma)} q_2$ iff
  $q_1\;\eqctx\leq_{\CtxA{\sigma(F)}}q_2\;\eqctx$;
  $\aeq[F'(\sigma)]{q}{e_1}{e_2}$ iff
  $\aeq[\CtxA{\sigma(F)}]{q\;\eqctx}{e_1\;\ectx}{e_2\;\ectx}$; and
  $e\in\appr[F'(\sigma)]{q}{a}$ iff
  $(e\;\ectx)\in\appr[\CtxA{\sigma(F)}]{q\;\eqctx}{a\;\actx}$.
  % \begin{itemize}
  % \item $q_1\leq_{F'(\sigma)} q_2$ iff $q_1\;\eqctx\leq_{\CtxA{\sigma(F)}}q_2\;\eqctx$;
  % \item $\aeq[F'(\sigma)]{q}{e_1}{e_2}$ iff $\aeq[\CtxA{\sigma(F)}]{q\;\eqctx}{e_1\;\ectx}{e_2\;\ectx}$; and
  % \item $e\in\appr[F'(\sigma)]{q}{a}$ iff $(e\;\ectx)\in\appr[\CtxA{\sigma(F)}]{q\;\eqctx}{a\;\actx}$.
  % \end{itemize}
  Intuitively, $\piappr{\apprctx}{\F}$ forms an approximation family
  where $\lamabs{\ectx}{e}$ is approximated by $\lamabs{\actx}{a}$ with
  error $\lamabs{\eqctx}{q}$ whenever $e$ is approximated by $a$ with
  error $q$ in approximation context $\apprctx$. When $\apprctx$ is just
  the single element $\aelemvar{x}{\F_1}$ and $\F_2$ does not depend on
  the values substituted for $\xe$, $\xa$, or $\xq$, then
  $\piappr{\apprctx}{\F_2}$, which we abbreviate as $\F_1\apprto\F_2$,
  yields the notion of function approximation types discussed in Section
  \ref{sec:overview} and in Example \ref{example:floating-point-functions}.
  % if $a_1$ is an approximation of $e_1$ with error
  % $q_1$ in $\F_1\apprto\F_2$, and if $a_2$ is an approximation under
  % $\F_1$ of exact input $e_2$ with error $q_2$, then the result
  % $a_1\;a_2$ is an approximation of the exact result $e_1\;e_2$ with
  % error $q_1\;e_2\;q_2$. Thus, $q_1$ essentially captures how the
  % error propagates through approximaton $a_1$, depending on the
  % input $a_2$ and its error.
\end{definition}

To approximate the polymorphic type $\foralltp{X}{E}$ we use
\[
\begin{array}{@{}l@{}}
\piappr{\atelemdef,
  \phielem{\zz\!:\!\Xq,\\\ind{1}\zp\!:\!\Xq\to\Xq\to\Xq}{
  \zz\ctxeq\zeroA[\xi]\wedge\zp\ctxeq\plusA[\xi]}}{\F}
\end{array}
\]
This approximation family quantifies over the type variables $\Xe$,
$\Xa$, and $\Xq$ for the exact, approximate, and error types, as well
as over the variables $\zz$ and $\zp$ for the zero error and error
addition of the approximation type $\xi$. The latter variables are
explicitly abstracted in order to allow error expressions to refer to
them: recall that $\xi$ is only bound in $\phictx$, not $\qctx$; i.e.,
error terms refer only to $\Xe$ and $\Xq$, not to $\xi$. Abusing
notation slightly, we abbreviate the above approximation family as
$\piapprtdef{\F}$.

\section{Verifying an Approximating Compiler}
\label{sec:approximating-compiler}

As discussed in the Introduction, the long-term goal of this work is
to enable language-based approximations, where a compiler or other
tool performs approximate transformations in a correct and automated
manner. In this section, we show how to verify such a tool, the goal
of the current work, with the semantics given in the previous
section. Specifically, we consider a tool that performs two
transformations: it compiles real numbers into floating-point
implementations; and it optionally performs loop perforation
\cite{misailovic11,sidiroglou11}. For the current work, we assume only
that our tool can be specified with an \emph{approximate compilation
  judgment} $\compilesto{e}{a}{q}{\A}$ such that this judgment can be
derived whenever the tool might approximate exact expression $e$ by
$a$ with error bound by $q$ using approximation type $\A$ and
assumptions $\apprctx$. Intuitively, each rule of this judgment
corresponds to an approximate transformation that the tool might
perform; for example, the approximation rule of Section
\ref{sec:overview}, for approximating $\sin$ by $\lamabs{x}{x}$, might
be included. We ignore the specifics of how the tool chooses which
approximations to use where, as long as all possible choices are
contained in the approximate compilation judgment.

To verify such a tool, we then prove soundness of its approximate
compilation judgment. Soundness here means that
$\compilesto{e}{a}{q}{\A}$ implies $e\in\appr[\CtxA{\A}]{q}{a}$. This
can be proved in a local, modular fashion, by verifying each
approximation rule individually; more specifically, if an
approximation rule derives $\compilesto{e}{a}{q}{\A}$ from assumptions
$\compilesto[\apprctx_i]{e_i}{a_i}{q_i}{\A_i}$ for $1\leq i\leq n$ and
side conditions $\phi_j$ for $1\leq j\leq m$, then the rule is correct
iff $e\in\appr[\CtxA{\A}]{q}{a}$ holds whenever
$e_i\in\appr[\CtxAprot{\apprctx_i}{\A_i}]{q_i}{a_i}$ for all $i$ and
$\phi_j$ holds for all $j$. This also allows extensibility, since
additional rules can always be added as long as they are proved
correct. In the remainder of this section, we consider rules that
would be used in our example tool, including: compositionality rules
(Section \ref{subsec:compositionality-rules}); rules for replacing
real numbers by floating-point implementations (Section
\ref{subsec:floating-point}); and a rule for performing loop
perforation (Section \ref{subsec:loop-perforation}).

%%%%%%%%%%%%%%%
% Sub-section %
%%%%%%%%%%%%%%%
\subsection{Compositionality Rules}
\label{subsec:compositionality-rules}

In order to combine errors from individual approximate transforms into
a single, whole-program error, we now introduce the
\emph{compositionality rules}. These rules, given in Figure
\ref{fig:compositionality}, are essentially the identity, stating that
each expresison construct can be approximated by itself; however, they
show how to build up and combine error expressions for different
contructs.

%%%%% Approximate Compilation %%%%%

% NOTE: we put this here to make it appear above the text describing it

\begin{figure*}[t]
  \centering
  \small
  \begin{tabular}{@{}c@{}}
    % error weakening
    \infer[\rulename{A-Weak}]{\compilesto{e}{a}{q'}{\F}}{\compilesto{e}{a}{q}{\F} & q\leq_{\CtxA{\F}} q'}
    \figfill

    % lambda abstractions
    \infer[\!\rulename{A-Lam}]{
      \compilesto{\lamabs{\xe}{e}}{\lamabs{\xa}{a}}{\lamabs{\xe}{\lamabs{\xq}{q}}}{\piapprdef{\F'}}
    }{
      \compilesto[\apprctx,\aelemabbr{x}{\F}]{e}{a}{q}{\F'}
    }
    \\ \\

    % \infer[\rulename{A-ReAppr}]{\compilesto{e}{a'\;a}{q'\;e\;q}{\F_2}}{\compilesto{e}{a}{q}{\F_1} \\ \lamabs{\xe}{\xe}\in\appr[\CtxA{\F_1\apprto\F_2}]{q'}{a'}}
    % \\[6pt]

    % error limits
%     \infer[\rulename{a-lim}]{\compilesto{e}{a}{q}{\F}}{\forall i.\compilesto{e}{a}{q_i}{\F} & \forall i.q_{i+1}\ltA q_i & \mathrm{glb}(\vec{q},q)}
%     \figfill

    % variables
    \infer[\rulename{A-Var}]{\compilesto{\xe}{\xa}{\xq}{\F}}{\aelemdef\in\apprctx}
    \figfill

    % % phi introduction
    % \infer[\rulename{A-Phi}]{\compilesto{e}{a}{(\lamabs{\Gamma}{q})}{\piappr{\phielemdef}{\F}}}{\compilesto[\apprctx,\phielemdef]{e}{a}{q}{\F}}
    % \\ \\

    % % phi elimination
    % \infer[\!\rulename{A-PhiE}]{
    %   \compilesto[\apprctx,\phielemdef]{e}{a}{q\;\vec{q}}{[\vec{q}/\Gamma]\F}
    % }{
    %   \compilesto{\!e\!}{\!a}{q}{\piappr{\phielemdef}{\F}} & \typejT{[\vec{q}/\Gamma]}{\phielemdef}
    % }
    % %\figfill

    % functional constants
    % \infer[\rulename{A-K}]{\compilesto{k}{a}{q}{\A}}{
    %   \quadtuple{k}{a}{q}{\A}\in\K
    % }
    % \\[6pt]

    % applications
    %\begin{tabular}{@{}c@{}}
    \infer[\!\rulename{A-App}]{
      \compilesto{e_1\;e_2}{a_1\;a_2}{q_1\;e_2\;q_2}{[e/\xe,a/\xa,q/\xq]\F'}
    }{
      \begin{array}{c}
        \compilesto{\!e_1\!}{\!a_1\!}{q_1\!}{\!\piapprdef{\F'}}
        %\\
        \compilesto{e_2\!}{\!a_2\!}{q_2}{\F}
        %\figfill
        %\Theta=[e_2/\xe,a_2/\xa,q_2/\xq]
      \end{array}
    }
    %\end{tabular}
    \\ \\

    % type lambdas
    \infer[\rulename{A-TLam}]{
       \compilestonl{\tplamabs{\Xe}{e}}{\tplamabs{\Xa}{a}}{\tplamabs{\Xe}{\tplamabs{\Xq}{\lamabs{\zz}{\lamabs{\zp}{q}}}}}{\piapprtdef{\F}}
    }{
      \compilesto[\apprctx,\atzelemabbr{X}{z}{\xi}]{e}{a}{q}{\F}
    }
    %\\ \\
    \figfill

    % type applications
    \infer[\!\rulename{A-TApp}]{
      \compilestonl{\tpapp{e}{\EF}}{\tpapp{a}{\AF}}{\tpapp{q}{\EF,\QQF}\;\zeroF\;\plusF}{[\F/\xi,\ldots]\F'}
      % \compilesto{\tpapp{e}{\EF}}{\tpapp{a}{\AF}}{\tpapp{q}{\EF,\QQF}\;\zeroF\;\plusF}{[\F/(\atzelemabbr{X}{z}{\xi})]\F'}
    }{
      \compilesto{e}{a}{q}{\piapprtdef{\F'}}
      &
      \typejF{\F}
    }
    \\ \\

    % functional constants (old version with \kapp command)
    % \infer[\rulename{A-K}]{\compilesto{\kapp{k}{\vec{\tau}}{\vec{e}}}{\tpapp{a}{\vec{\AFp}}\;\vec{a}}{\tpapp{q}{\overrightarrow{\EFp,\QQFp}}\;\overrightarrow{\zeroFp\;\plusFp}\;\overrightarrow{e\;q}}{\Theta(\F)}}{
    %   \figfill
    %   \quadtuple{k}{a}{q}{\piapprtmulti{\piapprmulti{\F}}}\in\K
    %   \figfill
    %   \forall i.\typejF{\F_i'}
    %   \figfill
    %   \\
    %   \Theta=[\vec{\F'}/(\atzelemabbr{\vec{X}}{\vec{z}}{\vec{\xi}}),(\vec{e},\vec{a},\vec{q})/\vec{x}]
    %   %
    %   \figfill
    %   \forall i.\compilesto{e_i}{a_i}{q_i}{\Theta(\F_i)}
    % }
    % \figfill[2pt]

    % fix expressions
    \infer[\rulename{A-Fix}]{
      \compilesto{\fixappnotp{e}}{\fixappnotp{e}}{\fixappnot{q\;({\fixappnotp{e}})}}{\F}
    }{
      \compilesto{e}{a}{q}{
          \piapprdef{\begin{array}[t]{@{}l@{}}(\xe\ctxeq\fixappnotp{e}\wedge\xa\ctxeq\fixappnotp{a}%\\
              \wedge\xq\ctxeq\fixappnot{q\;({\fixappnotp{e}})})\apprimp\F\end{array}}
%           \piapprdef{(\xe\ctxeq\fixappnop[\EF]{e}\wedge\xa\ctxeq\fixappnop[\AF]{a}\wedge\xq\leq\fixapp[\QQF]{q\;({\fixappnop[\EF]{e}})})\apprimp\F}
%         \begin{array}[t]{@{}l@{}}
%           \piapprdef{\\
%             \piappr{\phielem{\cdot}{\xq\leq\fixappnot{q\;({\fixappnot{e}})}}}{\F}}
%         \end{array}
      }}
    \\ \\

    % if expressions
    \infer[\rulename{A-If}]{
      \compilesto{\ifthenelse{e}{e_{\True}}{e_{\False}}}{\ifthenelse{a}{a_{\True}}{a_{\False}}}{q}{\F}
    }{
      \compilesto{e}{a}{q'}{\F'}
      &
      \forall b\in\{\True,\False\}.\compilesto{e_b}{a_b}{q_b}{\F}
      &
      \forall (b,b')\in\{\True,\False\}.
      e_{b}\in\appr[\CtxAprot{\apprctx,b\in\apprprot{\F'}{q'}{b'}}{\F}]{q}{a_{b'}}
    }

  \end{tabular}
  %\vspace{-0.05in}
  \caption{Compositionality Rules for Approximate Compilation}
  \label{fig:compositionality}
%\vspace{-0.15in}
\end{figure*}

The first rule, \rulename{A-Weak}, allows the error bound to be
weakened from $q$ to any greater error $q'$. The
\rulename{A-Var} rule approximates variable $\xe$ by $\xa$ with error
$\xq$ when these variables are associated in $\apprctx$.
% The \rulename{A-Phi} rule introduces a $\piappr{\phielemdef}{\F}$
% approximation by adding $\phielemdef$ to $\apprctx$.
% % Since the error $q$ can contain the variables in $\phielemdef$,
% % these variables are abstracted out of $q$ in the conclusion of the
% % rule using the notation $\lamabs{\Gamma}{q}$.
% The elimination rule for $\piappr{\phielemdef}{\F}$,
% \rulename{A-PhiE}, finds a sequence $\vec{q}$ of expressions to
% substitute for $\Gamma$ that satisfy $\phi$, performing this
% substitution into $\F$ and applying $q$ to $\vec{q}$ in the
% conclusion. For \rulename{A-K}, we assume as given a set \K\ of
% tuples $\quadtuple{k}{a}{q}{\A}$ for approximating each $k$ such
% that $k\in\appr[\A]{q}{a}$; the rule then returns one such $a$ and
% $q$.
The rule \rulename{A-Lam} approximates a lambda-abstraction
$\lamabs{\xe}{e}$ with a lambda-abstraction $\lamabs{\xa}{a}$ by
approximating the body $e$ by $a$, using the error function
$\lamabs{\xe}{\lamabs{\xq}{q}}$ that abstracts over the input $\xe$
and its approximation error $\xq$. The \rulename{A-App} rule
approximates applications $e_1\;e_2$ by applying the approximation of
$e_1$ to that of $e_2$ and applying the error for $e_1$ to both $e_2$
and its error. The \rulename{A-TLam} rule approximates polymorphic
lambdas $\tplamabs{\Xe}{e}$ by approximating the body $e$ in the
extended approximation context $\apprctx,\atzelemabbr{X}{z}{\xi}$,
recalling the abbreviation $\atzelemabbr{X}{z}{\xi}$ from Section
\ref{subsec:approximation-families} that abstracts the various
components of approximation families. \rulename{A-TApp} approximates
type applications $\tpapp{e}{\EF}$ where the type involved is the
exact type of some approximation family $\F$. This is accomplished by
first approximating $e$ to some $a$ with error $q$ in the polymorphic
approximation $\piapprtdef{\F'}$ introduced in Definition
\ref{def:pi-approximation}, and then applying $a$ to the approximate
type $\AF$ of $\F$. The error $q$ is applied to the necessary
components of $\F$, and the resulting approximation is $\F'$ with $\F$
substituted for $\xi$ and all the appropriate components of $\F$
substituted for the $X$ and $z$ variables. This is abbreviated as
$[\F/\xi,\ldots]\F'$.

Fixed-points $\fixappnot{e}$ are approximated using
\rulename{A-Fix}.
% noting that the type argument to \fixname\ will always be the same
% as the exact type $\EF$ of the current $\F$.
This approximates $e$ to $a$ with error $q$, applying $\fixname$ to
the results in the conclusion. The approximation family used for $e$ is
$\F\apprto\F$, augmented with the assumption that the inputs $\xe$,
$\xa$, and $\xq$ are equal to the exact, approximate, and error
\fixname-expressions in the conclusion of the rule.

Finally, if-expressions are approximated with \rulename{A-If}. First,
each component of the if-expression is approximated; the condition can
use an arbitrary approximation family $\F'$, while the then and else
branches must use the same family as the whole expression. The final
condition requires that the output error $q$ bounds the error between
the branches taken in the exact and approximate expressions, even if
one takes the then branch and the other takes the else branch. This is
stated by quantifying over all four combinations of $\True$ or
$\False$ in the exact and approximate conditions that meet the error
$q'$ computed for approximating the if-condition, and then requiring
that the corresponding then or else branches are within error $q$.

\begin{lemma}
  Each rule of Figure \ref{fig:compositionality} is sound.
\end{lemma}

%%%%%%%%%%%%%%%
% Sub-section %
%%%%%%%%%%%%%%%
\subsection{Floating-Point Approximation}
\label{subsec:floating-point}

To approximate real-number programs with their floating-point
equivalents, we use the $\FloatA$ approximation type formalized in
Example \ref{example:floating-point}. We then add rules
\[
\begin{array}{@{}c@{}}
\infer{\compilesto{r}{\mathsf{real}(r)}{|r-\mathsf{real}(r)|}{\FloatA}}{}
\\ \\
\infer{\compilesto{op^\R}{op^{\Float}}{op^{q}}{\FloatA\apprto\FloatA\apprto\FloatA}}{}
\end{array}
\]
for each real number $r$ and each built-in binary operation (such as
$+$, $*$, etc.) $op$, where $op^\R$ and $op^\Float$ are the version of
$op$ for reals and floating-points, respectively, and $op^{q}$ is
the error function calculating the size of the interval error in the
output from those of the inputs.  The error functions
$op^{q}$ to use for the various operations can be derived in a
straightforward manner by considering the smallest error that bounds
the difference between the real result of $op^r$ and any potential
result of $op^{\Float}$ on floating-point numbers in the input
interval. The infinite error $\bot$ is returned in case of
overflow or Not-A-Number results. For instance, the error
$+^{q}$ can defined (in pseudocode) as:
\[
\begin{array}{@{}l@{}}
  +^q\;xe\;\xq\;\ye\;\yq =\\
  \ind{1}
  \letin{I}{\mathsf{round}([(\xe-\xq)+(\ye-\yq),\\
    \hspace{78pt}(\xe+\xq)+(\ye+\yq)])}{\\
    \ind{2}\ifthenelse{\exists r\in I.|r| \geq \mathsf{MAXFLOAT}}{\bot}{\\
      \ind{3}\max_{r'\in I} |r'-r|
    }}
\end{array}
\]
where $\mathsf{round}$ rounds all reals in an interval to floating-point
numbers (using the current rounding mode) and $\mathsf{MAXFLOAT}$ is
the maximum absolute value of the floating-point representation
being used. The errors for other operations can be defined similarly.

Using these rules with the compositionality rules of Figure
\ref{fig:compositionality} yields an interval-based
floating-point error analysis that works for higher-order
and even polymorphic terms.
%  for example, the error for the
% polymorphic function
% \[
% \tplamabs{\Xe}{\lamabst{f}{\Xe\to\R}{\lamabst{\xe}{\Xe}{(f\;\xe)*(f\;\xe)}}}
% \]
% will, when applied to 
Although recent work has given more precise floating-point error
analyses than intervals \cite{goubault11,darulova11}, we
anticipate, as future work, that such approaches can also be
incorporated into our framework, allowing them to be used on
higher-order, polymorphic programs.

\begin{theorem}
  The rules listed above for compiling reals to floating-points are
  sound.
\end{theorem}

\subsection{Loop Perforation}
\label{subsec:loop-perforation}

% %%%%%%%%%%%%%%%
% % Sub-section %
% %%%%%%%%%%%%%%%
% \subsection{Loop Perforation and Limit Truncation}
% \label{subsec:approximate-optimizations}

% We now extend the approximate compilation rules to implement two
% powerful and non-trivial approximate transformations that reduce
% computation cost, given in Figure \ref{fig:optimizations}.
% The first is \emph{limit truncation}, where the exact computation
% computes the limit of a sequence and the approximate computation
% simply returns the $n$th element for some suitable $n$. To formalize
% this transformation, we first define the \emph{limit} of a ground
% expression $e$ of type $\N\to\EA$ for some ground approximation $\A$
% as follows. Let $e'$ satisfy the property that, for any
% $q>_{\A}\zeroA$, there is some $n$ such that, for all $m>n$, we have
% $\aeq[\A]{q}{e\;m}{e'}$. In this case, $e'$ is said to be a limit of
% $e$. Next, assume that our input language is augmented with a function
% symbol $\limA$ of type $(\N\to\EF)\to\EF$ such that $F_{\limA}(e)$
% returns a limit of $e$, if one exists, and returns $\bot$
% otherwise.

\begin{figure}[t]
  \centering
{\small 
  \begin{tabular}{@{}c@{}}

    % %% limit estimation
    % \infer[\rulename{A-Trunc}]{
    %   \compilesto{\limA\;e\!}{\!a\;\nna\!}{\!(q\;\nne\;\nnq)\plusA q'}{\A}
    % }{
    %   \begin{array}{@{}c@{}}
    %     \EF\equiv\N
    %     \figfill
    %     \compilesto{e}{a}{q}{\F\apprto\A}
    %     \figfill
    %     \compilesto{\nne}{\nna}{\nnq}{\F}
    %     % \nne\in\appr[\CtxAprot{\apprctx}{\F}]{\nnq}{\nna}
    %     % \hspace{-10pt}
    %     % \\
    %     \figfill
    %     \aeq[\CtxAprot{\apprctx}{\A}]{q'}{e\;\nne}{\limA\;e}
    %     % \figfill
    %   \end{array}
    % }
    % \\ \\

    %% fixed-point truncation: old e^\infty version
    % \infer[\rulename{A-Trunc}]{
    %   \compilesto{e^\infty\!}{\!a^{n}\!}{\!(q'\;\nne)\plusF(\lamabs{\pair{\xe}{\!\xq}}{\pair{e\;\xe}{\!q\;\xe\;\xq}})^{\nne}\!\!}{\F}
    % }{
    %   \compilesto{e}{a}{q}{\F\apprto\F}
    %   \figfill
    %   \nne\in\appr[\CtxAprot{\apprctx}{\ExactA{\N}}]{0}{n}
    %   \\
    %   \figfill
    %   \forall i.\aeq[\CtxAprot{\apprctx}{\F\apprto\F}]{q'\;i}{e^i}{e^\infty}
    %   \figfill
    % }
    % \\ \\

    %% reduce rule for loop perforation
    \infer%[\!\rulename{A-Perf}]
    {
      \compilestonlnl{\redseq\,e_1\,e_2\,e_3}{\redseq\,(\lamabs{x}{(a_1\,x)^K})\,( a_2/K)\,(\lamabs{x}{a_3(x*K)})}{
        (\redseq\plusF e_2\,(\lamabs{x}{(q_3\,\lfloor x\rfloor_{K}\,0)\plusF(q\,x)}))
        %\hspace{10pt}
        \!\plusF
        q'
      }{\F}
      %\hspace{-35pt}
    }{
      \begin{array}{@{}c@{}}
        \compilesto{e_1}{a_1}{q_1}{\F\apprto\F\apprto\F}
        \\
        \compilesto{e_2}{a_2}{0}{\NatA}
        \\
        \compilesto{e_3}{a_3}{q_3}{\NatA\apprto\F}
        \figfill
        \aeq[\CtxAprot{\apprctx,\ctxelem{x}{\N}}{\F}]{q\;x}{e_1\;x}{e_1\;\lfloor x\rfloor_{K}}
        \\
        \aeq[\CtxAprot{\apprctx}{\F}]{q'}{\redseq\;e_1\;e_2\;e_3}{\redseq\;e_1\;\lceil e_2\rceil_K\;e_3}
      \end{array}
    }

    %% old *list* perforation stuff
%     %% list perforation
%     \infer[\rulename{A-Perf}]{
%       \compilesto{e}{\tpapp{\perf}{\AF}\;a}{q+\perferr\;e}{\LperfA{\F}}
%     }{
%       \compilesto{e}{a}{q}{\LoneA{\F}}
%     }
%     \\ \\

%     %% map rule for list perforation
%     \infer[\rulename{A-Map}]{
%       \compilestonl{\tpapp{\map}{\EF[1],\EF[2]}\;e_1\;e_2}{\tpapp{\map}{\AF[1],\AF[2]}\;a_1\;a_2}{
%         \tpapp{\reduce}{\QQF[2]}\;\ub\;\zeroF\\
%         \;\;(\tpapp{\map}{\EF[1],\QQF[2]}\;(\lamabs{\xe}{q_1\;\xe\;q_2})\;(\tpapp{\perf}{\EF[1]}\;e_2))\\
%       }{\LperfA{\F_2}}
%       \hspace{-10pt}
%     }{
%       \begin{array}{@{}c@{}}
%         \compilesto{e_1}{a_1}{q_1}{\F_1\apprto\F_2}
%         \\
%         \compilesto{e_2}{a_2}{q_2}{\LperfA{\F_1}}
%       \end{array}
%     }
%     \\ \\

%     %% reduce rule for list perforation
%     \infer[\rulename{A-Red}]{
%       \compilestonl{\reduce\;e_1\;e_2\;e_3}{\reduce\;(\lamabs{x}{(a_1\;x)^2})\;a_2\;a_3}{
%         \tpapp{\folderrinf}{\EF[1],\QQF[1],\EF[2],\QQF[2]}\\
%         \;\;(\lamabs{\xe\!}{\!(e_1\;\xe)^2})
%         \;(\lamabs{\xe\!}{\!\lamabs{\xq\!}{(q_1\;\xq\;\xq)^2}})\\
%         \;\;e_2\;q_2\;(\perf\;e_2)\;q_3
%       }{\F_2}
%     }{
%       \begin{array}{@{}c@{}}
%         \compilesto{e_1}{a_1}{q_1}{\F_1\apprto\F_2\apprto\F_2}
%         \\
%         \compilesto{e_2}{a_2}{q_2}{\F_2}
%         \\
%         \compilesto{e_3}{a_3}{q_3}{\LperfA{\F_1}}
%       \end{array}
%     }

  \end{tabular}
}
\vspace{-0.05in}
%\caption{Rules for Fixed-Point Truncation and Loop Perforation}
\caption{Loop Perforation}
\label{fig:loop-perforation}
\vspace{-0.15in}
\end{figure}

% In general, limits are not computable.
% % This means that they cannot be
% % written directly in a traditional programming language. In an
% % approximate semantics, however, if we can prove that the $n$th element
% % is within distance $q$ of the limit, then we can approximate this
% % limit with the $n$th element of the sequence.  This in turn can
% % express, for example, $n$th-order Taylor or Fourier approximations.
% Limit truncation, given by the \rulename{A-Trunc} rule,
% approximates a limit $\limA\;e$ by first approximating
% $e$ by some $a$ with error $q$, and then finding a ``large enough''
% input $\nne$ such that $e\;\nne$ is within some distance $q'$ of
% $\limA\;e$. An approximation $\nna$ of $\nne$ is then passed to
% $a$, with the error combining the error of this function call
% with $q'$.

Loop perforation \cite{misailovic11,sidiroglou11} is a powerful
approximate transformation that takes a loop which combines $f(0)$,
$f(1)$, etc., for some computationally intensive function $f$, and
only performs every $n$th iteration, repeating the values that are
computed $n$ times. We formalize a simplified version of loop
perforation as operating on reductions $\redseq\;e_1\;e_2\;e_3$,
defined as the function that reduces, using $e_1$, the sequence of
values of $e_3$ from $0$ less than or equal to $e_2$. The rule
in Figure \ref{fig:loop-perforation} shows how to
perforate this reduction by first approximating each $e_i$ to some
$a_i$ with error $q_i$. The rule requires that the number $e_2$ of
sequence elements can be approximated with $0$ error; this is not a
fundamental limitation, but is simply made here to simplify the
exposition. The approximation type $\NatA$ is defined similarly to
$\FloatA$, except that the natural numbers are used for the exact,
approximate, and error types. Next, the rule finds an error function
$q$ that, when applied to input $x$, bounds the error between the
exact sequence value $e_1\;x$ and the $x$th value in the perforated
sequence, which can be calculated as %
$e_1\;\lfloor n\rfloor_K$. Finally, the rule considers the fact that
$n$ may not be an exact multiple of $K$, in which case the perforated
loop actually computes the result of $\lceil e_2\rceil_K$
iterations. An error $q'$ is thus synthesized to bound the error
between this perforated result and the original reduction.  The
resulting approximation then performs the perforation computation as
described above, and the error sums the errors %
$(q_3\,\lfloor x\rfloor_{K}\,0)$, capturing the approximation error of
$a_3$, with $(q\,x)$, capturing the error of using %
$\lfloor x\rfloor_{K}$ instead of $x$, for each value $x$, and then
adds the error $q'$.

% %%%%%%%%%%%%%%%
% % Sub-section %
% %%%%%%%%%%%%%%%
% \subsection{Fixed-Point Truncation}
% \label{subsec:fixed-point-truncation}

% %%%%%%%%%%%%%%%
% % Sub-section %
% %%%%%%%%%%%%%%%
% \subsection{List Perforation}
% \label{subsec:list-perforation}

% \begin{theorem}
%   The rule \rulename{A-Perf} is sound.
% \end{theorem}

\begin{theorem}
  The rule of Figure \ref{fig:loop-perforation} is sound.
\end{theorem}

%%%%%%%%%%%%%%%%%%%%%%%%%%%%%%%
%%%%%       Section       %%%%%
%%%%%%%%%%%%%%%%%%%%%%%%%%%%%%%
\section{Related Work}
\label{sec:related-work}

A number of recent papers relate to program
approximations. Possibly the closest to this work is the work of Reed
and Pierce \cite{reed10}, because they consider a higher-order (though
not polymorphic) input language.  They show how to perform an
approximate program transformation that adds noise to a database query
to ensure differential privacy, i.e., that a query cannot violate the
privacy of a single individual recorded in the database. In order to
ensure that adding noise does not change the query results too much, a
type system is used to capture functions that are $K$-Lipschitz,
meaning that a change of $\delta$ in the input yields a change of at
most $K*\delta$ in the output.  This condition can be captured in our
system by the error $\lamabs{\xe\!}{\!\lamabs{\xq\!}{K*\xq}}$ in an
approximation of a function.

Loop perforation \cite{misailovic11,sidiroglou11} transforms certain
map-reduce programs, written as \texttt{for}-loops in C, to perform
only a subset of their iterations. Section
\ref{subsec:loop-perforation} shows how to capture a variant
of this transformation in our system. More recent work \cite{zhu12}
has extended loop perforation to \emph{sampling}, which takes a random
(instead of a controlled) sample of the iterations of a loop. This
work also adds \emph{substitution} transformations, where different
implementations of the same basic operations (such as $\sin$ or
$\log$) are substituted to try to trade off accuracy for performance.

EnerJ \cite{sampson11} allows the programmer to specify, with type
modifiers, whether data is \emph{exact} or \emph{inexact}. Inexact
data means the program can tolerate errors in this data. Such data is
then stored in low-power memory, which is cheaper but is susceptible
to random bit flips.
% As discussed in Section \ref{subsec:approximation-examples}, these
% notions can be captures with the $\ExactA{\tau}$ and $\HomA{\tau}$
% approximations in our system.

Carbin et al.\ \cite{carbin12} present a programming model with
\emph{relaxation}, where certain values in a program are allowed to
vary in a pre-defined way, to model errors. The authors
then show how to verify properties of relaxed programs despite these
errors. Although this work addresses some of the same questions as
ours, one significant drawback is that it cannot represent the
relationship between expressions of different types, such as
real-valued functions and functions on floating-point numbers used to
implement them.

Approximate bisimulation \cite{girard11} is a new technique for
relating discrete, continuous and hybrid systems in a manner that can
tolerate errors in a system. We are still investigating the
relationships between our system and approximate bisimulation, but one
key difference with our work is that this approach considers only
transition systems, and so cannot be applied to higher-order programs,
programs with recursive data, etc.; however, such systems still
encompass the large and practical class of control systems.

Chaudhuri et al.\ \cite{chaudhuri11,chaudhuri10} have investigated a
static analysis for proving robustness of programs, which they then
argue is a useful precondition for approximate transformations. They
define robustness as the $K$-Lipschitz condition, which, again, can be
captured in our system as the error
$\lamabs{\xe\!}{\!\lamabs{\xq\!}{K*\xq}}$.
% There is also a strong
% similarity between the domain-specific reasoning in our
% \rulename{A-Case} rule and the boundary condition assumptions used in
% establishing robustness of \ifname-expressions, as mentioned
% in Section \ref{sec:applications}.

% FIXME: relate to worker/wrapper

Floating-point error analysis \cite{goubault11,darulova11} bounds the
error between a floating-point program and the syntacically equivalent
real-number program. We showed how to accomplish this in our setting
in Section \ref{subsec:floating-point}, yielding what we believe is
the first floating-point error analysis to work for higher-order and
polymorphic programs. Although state-of-the-art analyses use affine
expressions, instead of intervals as we used above, we anticipate that
we will be able to accommodate these approaches in our semantics as
well, thereby adapting them to higher-order and polymorphic programs.

On a technical note, our quantification types and approximate equality
relations were inspired by Flagg and Kopperman's continuity spaces
\cite{flagg97}, one of the few works that considers a more general
notion of distance than metric spaces. Specifically, the distributive
lattice completion of a quantification type corresponds exactly to
Flagg and Kopperman's abstract notion of distance, called a
\emph{quantale}.  A similar transformation turns an approximate
equality relation into a \emph{continuity space}, their generalization
of metric spaces.

%%%%%%%%%%%%%%%%%%%%%%%%%%%%%%%
%%%%%       Section       %%%%%
%%%%%%%%%%%%%%%%%%%%%%%%%%%%%%%
\section{Conclusion}
\label{sec:conclusion}

We have introduced a semantics for {\em approximate program
  transformations}. The semantics relates an \emph{exact} program to
an \emph{approximation} of it, and quantifies this relationship with
an \emph{error} expression.  Rather than specifying errors solely with
real numbers and metric spaces, our approach is based on
\emph{approximation types}, an extension of logical relations that
allows us to use, for example, functions as the errors for
approximations of functions, and polymorphic types as the errors for
polymorphic types. We then show how approximation types can be used to
verify approximate transforms in a modular, composable fashion, by
proving soundness of each transform individually and by including a
set of compositionality rules, also proved correct, that combine
errors from individual approximate transforms into a whole-program
error.
% Proving that an approximately transformed program satisfies user
% error bounds is then reduced to proving that this

% In ongoing work, we are extending the semantics presented here to
% approximate program transformations that are also {\em
%   randomized}~\cite{zhu12}, by using a probability monad \cite{RP02}
% to model probabilistic programs in our framework.
% % Also,
% % % this paper has focused on the semantics of errors introduced by
% % % approximate program transformations, rather than the {\em resource
% % %   savings} gained by these transformations. In current work, 
% %   we are currently extending our semantics with a quantification of
% %   the resources consumed by an approximation.
% We are also designing a high-level programming language based
% on our semantics to allow the compiler and runtime
% system to ``tune'' certain expressions in a program, causing the
% program to stay within a specified resource bound while satisfying an
% overall requirement of approximate correctness. Safety-critical
% embedded applications
% that function in a variety of resource-constrained environments
% % , but are at the same time safety-critical,
% seem to be a natural application domain for this language.

% \appendix
% \section{Appendix Title}

% This is the text of the appendix, if you need one.

% \acks

% Acknowledgments, if needed.

% We recommend abbrvnat bibliography style.

%\begin{spacing}{0.99}
%\bibliographystyle{abbrvnat}
\bibliographystyle{splncs03}
\bibliography{bib}
%\end{spacing}

\end{document}

%%% Local Variables: 
%%% mode: latex
%%% End: 